\documentclass[xelatex, 12pt]{article}

\usepackage{amsmath, amssymb, amsthm, amsfonts, bbm}
\usepackage[tbtags]{mathtools}
\usepackage{graphicx}
\usepackage{verbatim}
\usepackage{enumerate}
\usepackage{palatino}
\usepackage[usenames,dvipsnames]{color}
\usepackage{units}
\usepackage{tikz}
\usepackage{subfigure}
\usepackage{setspace}
\usepackage{enumitem}
\usepackage[utf8]{inputenc}
\usepackage[T1]{fontenc}
\usepackage[charter]{mathdesign}
\usepackage[top=1in, bottom=1in, left=1in, right=1in]{geometry}
\usepackage[pdfstartview=FitH, pdfpagemode=None, colorlinks=true, citecolor=blue, linkcolor=blue, urlcolor=magenta, pagebackref=true, unicode=false]{hyperref}

\usepackage{xargs} 
\usepackage[colorinlistoftodos,prependcaption,textsize=footnotesize]{todonotes}

\usetikzlibrary{positioning,snakes,mindmap,calc,fit,arrows,shapes}

\tikzstyle{box} = [shape=rectangle,draw=black,thick]

\newcommand{\myignore}[1]{}

\newcommand{\inbrace}[1]{\left \{ #1 \right \}}
\newcommand{\inparen}[1]{\left ( #1 \right )}
\newcommand{\insquare}[1]{\left [ #1 \right ]}
\newcommand{\inangle}[1]{\left \langle #1 \right \rangle}
\newcommand{\infork}[1]{\left \{ \begin{matrix} #1 \end{matrix} \right .}
\newcommand{\inmat}[1]{\begin{matrix} #1 \end{matrix}}

\newcommand{\set}[1]{\inbrace{#1}}
\newcommand{\setdef}[2]{\set{#1 : #2}}
\newcommand{\defeq}{\stackrel{\mathrm{def}}{=}}

\newcommand{\sAND}{\ \mathrm{and} \ }

\newcommand{\union}{\cup}

\newcommand{\bit}{\set{0,1}}

\newcommand{\wtilde}[1]{\widetilde{#1}}

\newcommand{\bbZ}{{\mathbb Z}}

\newcommand{\bbN}{{\mathbb N}}

\newcommand{\ba}{\mathbf{a}}

\newcommand{\be}{\mathbf{e}}

\newcommand{\bu}{\mathbf{u}}
\newcommand{\bv}{\mathbf{v}}

\newcommand{\bx}{\mathbf{x}}
\newcommand{\by}{\mathbf{y}}

\newcommand{\bX}{\mathbf{X}}
\newcommand{\bY}{\mathbf{Y}}

\newcommand{\calJ}{\mathcal{J}}

\newcommand{\poly}{\mathrm{poly}}
\newcommand{\supp}{\mathrm{supp}}

\newcommand{\IC}{\mathrm{IC}}
\newcommand{\CC}{\mathrm{CC}}

\newtheorem{theorem}{Theorem}[section]
\newtheorem{defn}[theorem]{Definition}

\newtheorem{lem}[theorem]{Lemma}
\newtheorem{proposition}[theorem]{Proposition}
\newtheorem{cor}[theorem]{Corollary}

\newtheorem{observation}[theorem]{Observation}
\newtheorem{ex}[theorem]{Exercise}

\newtheorem{note}[theorem]{Note}

\newtheorem{remark}[theorem]{Remark}

\newenvironment{proof-overview}{\noindent {\em Proof Overview.} \hspace*{1mm}}{\hfill $\Box$}
\newenvironment{proofof}[1]{\noindent {\em Proof of #1.} \hspace*{1mm}}{\hfill $\Box$}
\newenvironment{proof-sketch}{\noindent {\em Sketch of Proof.} \hspace*{1mm}}{\hfill $\Box$}
\newenvironment{proof-idea}{\noindent{\em Proof Idea.}\hspace*{1mm}}{\qed\bigskip}
\newenvironment{proof-attempt}{\noindent{\em Proof Attempt.}\hspace*{1mm}}{\qed\bigskip}

\newcommand{\R}{\mathrm{R}}

\newcommand{\oneway}{\text{$1$-way}}

\newcommand{\PSRCC}{\text{PSR-CC}}

\newcommand{\GHD}{\mathrm{GHD}}

\newcommand{\UDISJ}{\text{\sc UDisj}}
\newcommand{\SI}{\text{\sc SparseInd}}
\newcommand{\SSI}{\mathrm{SSI}}
\newcommand{\LUDISJ}{\text{\sc LUDisj}}
\newcommand{\eGHD}{\overline{\GHD}}
\newcommand{\HD}{\mathrm{HD}}

\newcommand{\DSBS}{\mathrm{DSBS}}

\newcommand{\setinc}{\text{\sc SetInc}}
\renewcommand{\mod}{\mathrm{mod}\ }
\newcommand{\Otilde}{\widetilde{O}}
\newcommand{\bits}{\set{0,1}}

\newcommand{\isr}{\mathrm{ISR}}

\def \focs{0}

\def \todotoggle{0}
\newcommand{\mytodo}[1]{\ifnum\todotoggle=1{#1}\fi}

\newcommandx{\unsure}[2][1=]{\mytodo{\todo[linecolor=blue,backgroundcolor=blue!25,bordercolor=blue,#1]{#2}}}
\newcommandx{\change}[2][1=]{\mytodo{\todo[linecolor=cyan,backgroundcolor=cyan!25,bordercolor=cyan,#1]{#2}}}
\newcommandx{\info}[2][1=]{\mytodo{\todo[linecolor=green,backgroundcolor=green!25,bordercolor=green,#1]{#2}}}
\newcommandx{\improve}[2][1=]{\mytodo{\todo[linecolor=red,backgroundcolor=red!25,bordercolor=red,#1]{#2}}}
\newcommandx{\thiswillnotshow}[2][1=]{\todo[disable,#1]{#2}}
\newcommand{\tableoftodos}{\ifnum\todotoggle=1 \listoftodos[Comments/To Do's] \fi}

\usepackage{thmtools}
\usepackage{thm-restate}

\begin{document}

\title{Communication Complexity of Permutation-Invariant Functions}
\author{%
Badih Ghazi\thanks{Computer Science and Artificial Intelligence Laboratory, Massachusetts Institute of Technology, Cambridge MA 02139. Supported in part by NSF STC Award CCF 0939370 and NSF Award CCF-1217423.  {\tt badih@mit.edu}.} 
\and 
Pritish Kamath\thanks{Computer Science and Artificial Intelligence Laboratory, Massachusetts Institute of Technology, Cambridge MA 02139. Supported by NSF CCF-1420956.  {\tt pritish@mit.edu}.} 
\and 
Madhu Sudan\thanks{Microsoft Research, One Memorial Drive, Cambridge, MA 02142, USA. {\tt madhu@mit.edu}.}%
}
\date{\today}

\maketitle

\abstract{
Motivated by the quest for a broader understanding of communication complexity of simple functions, we introduce the class of ``permutation-invariant'' functions. A partial function $f:\bit^n \times \bit^n\to \set{0,1,?}$ is permutation-invariant if for every bijection $\pi:\set{1,\ldots,n} \to \set{1,\ldots,n}$ and every $\bx, \by \in \bit^n$, it is the case that $f(\bx, \by) = f(\bx^{\pi}, \by^{\pi})$. Most of the commonly studied functions in communication complexity are permutation-invariant. For such functions, we present a simple complexity measure (computable in time polynomial in $n$ given an implicit description of $f$) that describes their communication complexity up to polynomial factors and up to an additive error that is logarithmic in the input size. This gives a coarse taxonomy of the communication complexity of simple functions. Our work highlights the role of the well-known lower bounds of functions such as {\sc Set-Disjointness} and {\sc Indexing}, while complementing them with the relatively lesser-known upper bounds for {\sc Gap-Inner-Product} (from the sketching literature) and {\sc Sparse-Gap-Inner-Product} (from the recent work of Canonne et al. [ITCS 2015]). We also present consequences to the study of communication complexity with imperfectly shared randomness where we show that for total permutation-invariant functions, imperfectly shared randomness results in only a polynomial blow-up in communication complexity after an additive $O(\log \log n)$ overhead.
}

\newpage

\tableofcontents

\tableoftodos

\newpage

\section{Introduction} \label{sec:intro}

Communication complexity, introduced by Yao~\cite{yao1979some}, has been  a central object of study in complexity theory. In the {\em two-way model}, two players, Alice and Bob, are given private inputs $\bx$ and $\by$ respectively, along with some shared randomness, and they exchange bits according to a predetermined protocol and produce an output. The protocol computes a function $f(\cdot,\cdot)$ if the output equals $f(\bx,\by)$ with high probability over the randomness \footnote{In this work, we also consider partial functions where $f(\bx,\by)$ may sometimes be undetermined, denoted $f(\bx,\by) =~?$. In such cases, the protocol can output anything when $f(\bx,\by)$ is undetermined.}. The communication complexity of $f$ is the minimum over all protocols computing $f$ of the maximum, over inputs $\bx$ and $\by$, of the  number of bits exchanged by the protocol. The one-way communication model is defined similarly except that all the communication is from Alice to Bob and the output is produced by Bob. For an overview of communication complexity, we refer the reader to the book \cite{kushilevitz_nisan} and the survey \cite{lee2009lower}.

While communication complexity of functions has been extensively studied, the focus typically is on lower bounds. Lower bounds on communication complexity turn into lower bounds on Turing machine complexity, circuit depth, data structures, streaming complexity, just to name a few. On the other hand, communication complexity is a very natural notion to study on its own merits and indeed positive results in communication complexity can probably be very useful in their own rights, by suggesting efficient communication mechanisms and paradigms in specific settings. For this perspective to be successful, it would be good to have a compact picture of the various communication protocols that are available, or even the ability to determine, given a function $f$, the best, or even a good, communication protocol for $f$. Of course such a goal is overly ambitious. For example, the seminal work of Karchmer and Wigderson~\cite{karchmer1990monotone} implies that finding the best protocol for $f$ is as hard as finding the best (shallowest) circuit for some related function $\tilde{f}$.

Given this general barrier, one way to make progress is to find a restrictive, but natural, subclass of all functions and to characterize the complexity of all functions within this class. Such approaches have been very successful in the context of non-deterministic computation by restricting to satisfiability problems~\cite{Schaefer}, in optimization and approximation by restricting to constraint satisfaction problems~\cite{Creignou,KSTW}, in the study of decision tree complexity by restricting to graph properties~\cite{AKR}, or in the study of property testing by restricting to certain symmetric properties (see the survey~\cite{Sudan-Survey,goldreich2010property} and the references therein). In the above cases, the restrictions have led to characterizations (or conjectured characterizations) of the complexity of all functions in the restricted class. In this work, we attempt to bring in a similar element of unification to communication complexity.

In this work, we introduce the class of ``permutation-invariant'' (total or partial) functions. Let $[n]$ denote the set $\set{1, \cdots ,n}$. A function $f:\bit^n \times \bit^n\to \set{0,1,?}$ is permutation invariant if for every bijection $\pi:[n]\to[n]$ and every $\bx, \by \in \bit^n$ it is the case that $f(\bx, \by) = f(\bx^{\pi}, \by^{\pi})$. We propose to study the communication complexity of this class of functions.

To motivate this class, we note that most of the commonly studied functions in communication complexity including {\sc Equality}~\cite{yao1979some}, {\sc (Gap) Hamming distance} \cite{woodruff2004optimal, jayram2008one, chakrabarti2012optimal, vidick2011concentration, sherstov2012communication, pang1986communication, yao2003power, huang2006communication, blais_et_al:LIPIcs:2014:4717}, {\sc (Gap) Inner Product}, {\sc (Small-Set) Disjointness} \cite{kalyanasundaram1992probabilistic, razborov1992distributional, haastad2007randomized, saglam2013communication}, {\sc Small-Set Intersection} \cite{brody2014beyond} are all permutation-invariant functions. Other functions, such as {\sc Indexing} \cite{jayram2008one}, can be expressed without changing the input length significantly, as permutation-invariant functions.
Permutation-invariant functions also include as subclasses, several classes of functions that have been well-studied in communication complexity, such as (AND)-symmetric functions \cite{buhrman2001communication, razborov2003quantum,sherstov2011unbounded} and XOR-symmetric functions \cite{zhang2009communication}. It is worth noting that permutation-invariant functions are completely expressive if one allows an exponential blow-up in input size, namely, for every function $f(\bx,\by)$ there are functions $F$, $A$, $B$ s.t. $F$ is permutation-invariant and $f(\bx, \by) = F(A(\bx), B(\by))$. So results on permutation-invariant functions that don't depend on the input size apply to all functions. Finally, we point out that permutation-invariant functions have an important standpoint among functions with small communication complexity, as permutation-invariance often allows the use of hashing/bucketing based strategies, which would allow us to get rid of the dependence of the communication complexity on the input length $n$. We also note that functions on non-Boolean domains that are studied in the literature on sketching such as distance estimation (given $\bx,\by \in \R^n$, decide if $\|\bx - \by\|_p \leq d$ or if $\|\bx - \by\|_p > d(1+\epsilon)$) are also permutation-invariant. In particular, the resulting sketching/communication protocols are relevant to (some functions in) our class.

\subsection{Coarse characterization of Communication Complexity}
Permutation-invariant functions on $n$-bits are naturally succinctly described (by $O(n^3)$ bits). Given this natural description, we introduce a simple combinatorial measure $m(f)$ (which is easy to compute, in particular in time $\poly(n)$ given $f$) which produces a coarse approximation of the communication complexity of $f$. We note that our objective is different from that of the standard objectives in the study of communication complexity lower bounds, where the goal is often to come up with a measure that has nice mathematical properties, but may actually be more complex to compute than communication complexity itself. In particular, this is true of the Information Complexity measure introduced by \cite{chakrabarti2001informational} and \cite{bar2002information}, and used extensively in recent works, and which, until recently, was not even known to be approximately computable \cite{braverman2015information} (whereas communication complexity can be computed exactly, albeit in doubly exponential time). Nevertheless, our work does rely on known bounds on the information complexity of some well-studied functions and our combinatorial measure $m(f)$ also coarsely approximates the information complexity for all the functions that we study. 

To formally state our first theorem, let $\R(f)$ denote the randomized communication complexity of a function $f$ and $\IC(f)$ denote its information complexity. Our result about our combinatorial measure $m(f)$ (see Definition~\ref{def:measure_2way_partial}) is summarized below.

\begin{restatable}{theorem}{mainthmone}\label{thm:main-1}
Let $f:\{0,1\}^n \times \{0,1\}^n \to \{0,1,?\}$ 
be a (total or partial) permutation-invariant function.  Then, 
$$m(f) \leq \IC(f) \leq \R(f) \le \poly(m(f)) + O(\log{n}).$$
\end{restatable}


In other words, the combinatorial measure $m(f)$ approximates communication complexity to within a polynomial factor, up to an additive $O(\log n)$ factor. Our result is constructive --- given $f$ it gives a communication protocol whose complexity is bounded from above by $\poly(m(f)) + O(\log n)$. It would be desirable to get rid of the $O(\log n)$ factor but this seems
hard without improving the state of the art vis-a-vis communication complexity and information complexity. To see this, first note that our result above also implies that information complexity 
provides a coarse approximator for communication complexity. Furthermore, any improvement to the additive $O(\log n)$ error in this relationship would imply improved relationship between information complexity and communication complexity for general functions (better than what is currently known). Specifically, we note:

\begin{proposition}
Let $G(\cdot)$ be a function such that 
$\R(f) \le \poly(\IC(f)) + G(\log{n})$ for every permutation-invariant
partial function $f$ on $\bit^n\times \bit^n$.
Then, for every (general) partial function $g$ on $\{0,1\}^n \times \{0,1\}^n$,  
we have $\R(g) \le \poly(\IC(g)) + G(n)$.
\end{proposition}

Thus, even an improvement from an additive $O(\log n)$ to additive
$o(\log n)$ would imply new relationships between information
complexity and communication complexity.

\subsection{Communication with imperfectly shared randomness}

Next, we turn to communication complexity when the players only share randomness imperfectly, a model introduced by \cite{bavarian2014role,CGMS_ISR}. Specifically, we consider the setting where Alice gets a sequence of bits $r = (r_1,\ldots,r_t)$ and Bob gets a sequence of bits $s = (s_1,\ldots,s_t)$ where the pairs $(r_i,s_i)$ are identically and independently distributed according to distribution $\DSBS(\rho)$, which means, the marginals of $r_i$ and $s_i$ are uniformly distributed in $\bit$ and $r_i$ and $s_i$ are $\rho$-correlated (i.e., $\Pr[r_i = s_i] = 1/2 +\rho/2$). 

The question of what can interacting players do with such a correlation has been investigated in many different contexts including information theory~\cite{gacs1973common,witsenhausen1975sequences}, probability theory~\cite{mossel2005coin,bogdanov2011extracting,chan2014extracting,mossel2006non}, cryptography~\cite{brassard1994secret,maurer1993secret,renner2005simple} and quantum computing~\cite{bennett1996purification}. In the context of communication complexity, however, this question has only been investigated recently. In particular, Bavarian et al.~\cite{bavarian2014role} study the problem in the Simultaneous Message Passing (SMP) model and Canonne et al.~\cite{CGMS_ISR} study it in the standard one-way and two-way communication models. Let $\isr_\rho(f)$ denote the communication complexity of a function $f$ when Alice and Bob have access to $\rho$-correlated bits. The work of~\cite{CGMS_ISR} shows that for any total or partial function $f$ with communication complexity $R(f) \leq k$ it is the case that $\isr_\rho(f) \leq \min\{O(2^k), k + O(\log{n})\}$. They also give a partial function $f$ with $R(f) \leq k$ for which $\isr_\rho(f) = \Omega(2^k)$. Thus, imperfect sharing leads to an exponential slowdown for low-communication promise problems.

One of the motivations of this work is to determine if the above result is tight for total functions. Indeed, for most of the common candidate functions with low-communication complexity such as {\sc Small-Set-Intersection} and {\sc Small-Hamming-Distance}, we show (in Section~\ref{sec:atomic_isr_ub}) that $\isr_\rho(f) \leq \poly(\R(f))$. \footnote{In fact, this polynomial relationship holds for a broad subclass of permutation-invariant functions that we call ``strongly permutation-invariant''. A function $f(x,y)$ is strongly permutation invariant if there exists $h:\{0,1\}^2\to\{0,1\}$ and symmetric function $\sigma:\{0,1\}^n \to \{0,1\}$ such that $f(x,y) = \sigma(h(x_1,y_1),\ldots,h(x_n,y_n))$. Theorem~\ref{thm:strong_pi} shows a polynomial relationship between $\R(f)$ and $\isr(f)$ for all strongly-permutation-invariant total functions $f$.} This motivates us to study the question more systematically and we do so by considering permutation-invariant total functions. For this class, we show that the communication complexity with imperfectly shared randomness is within a polynomial of the communication complexity with perfectly shared randomness up to an additive $O(\log \log n)$ factor; this is a tighter connection than what is known for general functions. Interestingly, we achieve this by showing that the same combinatorial measure $m(f)$ also coarsely captures the communication complexity under imperfectly shared randomness. Once again, we note that the $O(\log \log n)$ factor is tight unless we can improve the upper bound of \cite{CGMS_ISR}.

\begin{restatable}{theorem}{mainthmtwo}\label{thm:main-2+3}
Let $f:\bit^n\times \bit^n \to \bit$ be a permutation-invariant total function. Then, we have
$$\isr_\rho(f) \leq \poly(\R(f)) + O(\log\log n)$$
Furthermore, $\isr^{\oneway}_\rho(f) \leq \poly(\R^{\oneway}(f)) + O(\log\log n)$.
\end{restatable}


\subsection{Overview of Proofs}

Our proof of Theorem~\ref{thm:main-1} starts with the simple observation
that for any permutation-invariant partial function $f(\cdot,\cdot)$, its
value $f(\bx,\by)$ is determined completely by $|\bx|$, $|\by|$ and $\Delta(\bx,\by)$
(where $|\bx|$ denotes the Hamming weight of $\bx$ and $\Delta(\bx,\by)$ denotes the
(non-normalized) Hamming distance between $\bx$ and $\by$). By letting Alice
and Bob exchange $|\bx|$ and $|\by|$ (using $O(\log n)$ bits of communication),
the problem now reduces to being a function only of the Hamming distance
$\Delta(\bx,\by)$. To understand the remaining task, we introduce a
multiparameter version of the Hamming distance problem
$\GHD_{a,b,c,g}^n(\cdot,\cdot)$ where $\GHD_{a,b,c,g}^n(\bx,\by)$ is undefined
if $|\bx| \ne a$ or $|\by| \ne b$ or $c-g < \Delta(\bx,\by) < c+g$. The function is
$1$ if $\Delta(\bx,\by) \geq c+g$ and $0$ if $\Delta(\bx,\by) \leq c-g$.

This problem turns out to have different facets for different choices of the
parameters. For instance, if $a \approx b \approx c$, then the communication complexity of
this problem is roughly $O((c/g)^2)$ and the optimal lower bound follows from the
lower bound on Gap Hamming Distance~\cite{chakrabarti2012optimal, vidick2011concentration, sherstov2012communication} whereas the upper bound
follows from simple hashing. However, when $a \ll b$, 
$c \approx b$ and $g \approx a$
different bounds and protocols kick in. In this range, the communication
complexity turns out to be $O(\log (c/g))$ with the upper bound coming from
the protocol for Sparse-Gap-Inner-Product given in \cite{CGMS_ISR}, and a
lower bound that we give based on a reduction from Set Disjointness.
In this work, we start by giving a complete picture of the complexity of
$\GHD$ for all parameter settings. The lower bound for communication
complexity, and even information complexity, of general
permutation-invariant functions $f$ follows immediately - we just look for
the best choice of parameters of $\GHD$ that can be found in $f$. The upper
bound requires more work in order to ensure that Alice and Bob can quickly narrow
down the Hamming distance $\Delta(\bx,\by)$ to a range where the value of $f$ is
clear. To do this, we need to verify that $f$ does not change values \emph{too
quickly} or \emph{too often}. The former follows from the fact that hard 
instances of $\GHD$ cannot be embedded in $f$, and the latter involves some
careful accounting, leading to a full resolution.

Turning to the study of communication with imperfectly shared randomness, we
hit an immediate obstacle when extending the above strategy since Alice
and Bob cannot
afford to exchange $|\bx|$ and $|\by|$ anymore, since this would involve $\Omega(\log
n)$ bits of communication and we only have an additional budget of $O(\log
\log n)$. Instead, we undertake a partitioning of the ``weight-space'',
i.e., the set of pairs $(|\bx|,|\by|)$, into a finite number of regions.
For most of the regions, we reduce the communication task to one of
the {\sc Small-Set-Intersection} or {\sc Small-Hamming-Distance} problems.
In the former case, the sizes of the sets are polynomially related to the randomized communication complexity, whereas in the latter case, the Hamming distance threshold is polynomially related to the communication complexity.
A naive conversion to protocols for the imperfectly shared setting using the results of \cite{CGMS_ISR} would result in an
exponential blow-up in the communication complexity. 
We give new protocols with imperfectly shared randomness for these two
problems
(which may be viewed as extensions of protocols in \cite{bavarian2014role} and
\cite{CGMS_ISR}) that manage to reduce the communication blow-up to just a
polynomial. This manages to take care of most regions, but not all.
To see this, note that any total function $h(|\bx|,|\by|)$ can be encoded as a
permutation-invariant function $f(\bx,\by)$ and such functions cannot be
partitioned into few classes. Our classification manages to eliminate
all cases except such functions, and in this case, we apply Newman's theorem
to conclude that the randomness needed in the perfectly shared setting is
only $O(\log \log n)$ bits (since the inputs to $h$ are in the range
$[n]\times[n]$). Communicating this randomness and then executing the
protocol with perfectly shared randomness gives us in this case a private-randomness
protocol with communication $\R(f) + O(\log \log n)$.

\subsection{Roadmap of this paper}
In Section~\ref{sec:prelim}, we give some of the basic
definitions and introduce the background material necessary for understanding the contributions of this paper. In Section~\ref{sec:ic_vs_cc}, we introduce our measure $m(f)$ and prove Theorem~\ref{thm:main-1}. In Section~\ref{sec:isr}, we show the connections between communication complexity with imperfectly shared randomness and that with perfectly shared randomness and prove Theorem~\ref{thm:main-2+3}. We end with a summary and some future directions in Section~\ref{sec:conclusion}.


\section{Preliminaries} \label{sec:prelim}

In this section, we provide all the necessary background needed to understand the contributions in this paper. 

\subsection{Notations and Definitions}

Throughout this paper, we will use bold letters such as $\bx$, $\by$, etc. to denote strings in $\bit^n$, where the $i$-th bit of $\bx$ will be accessed as $x_i$. We denote by $|\bx|$ the \emph{Hamming weight} of binary string $\bx$, i.e., the number of non-zero coordinates of $\bx$. We will also denote by $\Delta(\bx,\by)$ the \emph{Hamming distance} between binary strings $\bx$ and $\by$, i.e., the number of coordinates in which $\bx$ and $\by$ differ. We also denote $[n] \defeq \set{1,\cdots,n}$ for every positive integer $n$.

Very significant for our body of work is the definition of {\em permutation-invariant} functions, which we define as follows,

\begin{defn}[Permutation-Invariant functions]
A (total or partial) function $f : \bit^n \times \bit^n \to \set{0, 1, ?}$ is permutation-invariant if for all $\bx, \by \in \set{0, 1}^n$ and every bijection $\pi:[n] \to [n]$, $f(\bx^\pi, \by^\pi) = f(\bx, \by)$ (where $\bx^\pi$ is such that $x^{\pi}_i = x_{\pi(i)}$).
\end{defn}

\ifnum\focs=1

We let $\R(f)$ denote the probabilistic communication complexity of a function $f$ with shared randomness and error $1/$. We let $\isr_\rho(f)$ to denote the communication complexity of $f$ when Alice and Bob have access to a sequence of $\rho$-correlated bits~\cite{CGMS_ISR}. (We suppress the subscript if $0 < \rho < 1$.) We use $\IC(f)$ to denote the (prior-free) Interactive Information Complexity of $f$~\cite{braverman2012interactive}.

\else

\noindent We note the following simple observation about permutation-invariant functions.

\begin{observation}\label{obs:interchang_rep}
Any permutation-invariant function $f$ depends only on $|\bx \wedge \by|$, $|\bx \wedge \lnot \by|$, $|\lnot \bx \wedge \by|$ and $|\lnot \bx \wedge \lnot \by|$. Since these numbers add up to $n$, $f$ really depends on any three of them, or in fact any three linearly independent combinations of them. Thus, we have that for some appropriate functions $g$, $h$,
$$f(\bx, \by) = g(|\bx|, |\by|, |\bx \land \by|) = h(|\bx|, |\by|, \Delta(\bx, \by))$$
We will use these $3$ representations of $f$ interchangeably throughout this paper. We will often refer to the {\em slices} of $f$ obtained by fixing $|\bx| = a$ and $|\by| = b$ for some $a$ and $b$, in which case we will denote the sliced $h$ by either $h_{a,b}(\cdot)$ or $h(a,b,\cdot)$, and similarly for $g$.
\end{observation}

\subsection{Communication Complexity}

We define the standard notions of two-way (resp. one-way) randomized commmunication complexity\footnote{we will often abbreviate ``communication complexity'' by $\CC$} $R(f)$ (resp. $\R^{\oneway}(f)$), that is studied under shared/public randomness model (cf. \cite{kushilevitz_nisan}).

\begin{defn}[Randomized communication complexity $\R(f)$] For any function $f : \bit^n \times \bit^n \to \set{0,1,?}$, the randomized communication complexity $\R(f)$ is defined as the cost of the smallest randomized protocol, which has access to public randomness, that computes $f$ correctly on any input with probability at least $2/3$. In particular,
$$\R(f) \quad = \quad \min_{\substack{\Pi \\ \forall \bx, \by \in \bit^n \ s.t. \ f(\bx, \by) \ne \ ? :\\ \Pr[\Pi(\bx,\by) = f(\bx,\by)] \ge 2/3}} \ \CC(\Pi)$$
where the minimum is taken over all randomized protocols $\Pi$, where Alice and Bob have access to public randomness.

The {\em one-way randomized communication complexity} $\R^{\oneway}(f)$ is defined similarly, with the only difference being that we allow only protocols $\Pi$ where only Alice communicates to Bob, but not other way round.
\end{defn}

Another notion of randomized communication complexity that is studied, is under private randomness model. The work of \cite{CGMS_ISR} sought out to study an intermediate model, where the two parties have access to i.i.d. samples from a correlated random source $\mu(r,s)$, that is, Alice has access to $r$ and Bob has access to $s$. In their work, they considered the {\em doubly symmetric binary source}, parametrized by $\rho$, defined as follows,

\begin{defn}[Doubly Symmetric Binary Source $\DSBS(\rho)$] 
$\DSBS(\rho)$ is a distribution on $\bit \times \bit$, such that for $(r,s) \sim \DSBS(\rho)$,
$$\Pr[r=1,\, s=1] = \Pr[r=0,\, s=0] = (1+\rho)/4$$
$$\Pr[r=1,\, s=0] = \Pr[r=0,\, s=1] = (1-\rho)/4$$
\end{defn}

Note that $\rho = 1$ corresponds to the standard notion of public randomness, and $\rho = 0$ corresponds to the standard notion of private randomness.

\begin{defn}[Communication complexity with imperfectly shared randomness \cite{CGMS_ISR}]
For any function $f : \bit^n \times \bit^n \to \set{0,1,?}$, the ISR-communication complexity $\isr_\rho(f)$ is defined as the cost of the smallest randomized protocol, where Alice and Bob have access to samples from $\DSBS(\rho)$, that computes $f$ correctly on any input with probability at least $2/3$. In particular,
$$\isr_\rho(f) \quad = \quad \min_{\substack{\Pi \\ \forall \bx, \by \in \bit^n \ s.t. \ f(\bx, \by) \ne \ ? :\\ \Pr[\Pi(\bx,\by) = f(\bx,\by)] \ge 2/3}} \ \CC(\Pi)$$
where the minimum is taken over all randomized protocols $\Pi$, where Alice and Bob have access to samples from $\DSBS(\rho)$.

\noindent For ease of notation, we will often drop the subscript $\rho$ and denote $\isr(f) \defeq \isr_{\rho}(f)$.
\end{defn}

We use the term ISR as abbreviation for ``Imperfectly Shared Randomness'' and ISR-CC for ``ISR-Communication Complexity''. To emphasize the contrast, we will use PSR and PSR-CC for the classical case of (perfectly) shared randomness. It is clear that if $\rho > \rho'$, then $\isr_{\rho}(f) \le \isr_{\rho'}(f)$.

An extreme case of ISR is when $\rho = 0$. This corresponds to communication complexity with {\em private} randomness, denoted by $\R_{\mathrm{priv}}(f)$. Note that $\isr_{\rho}(f) \le \R_{\mathrm{priv}}(f)$ for any $\rho > 0$. A theorem (due to Newman \cite{newman1991private}) shows that any communication protocol using public randomness can be simulated using only private randomness with an extra communication of additive $O(\log n)$ (both in the 1-way and 2-way models). We state the theorem here for the convenience of the reader.

\begin{theorem}[Newman's theorem \cite{newman1991private}] \label{thm:newman}
For any function $f : \bit^n \times \bit^n \to \Omega$ (any range $\Omega$), the following hold,
\begin{eqnarray*}
\R_{\mathrm{priv}}(f) &\le& \R(f) + O(\log n)\\
\R_{\mathrm{priv}}^{\oneway}(f) &\le& \R^{\oneway}(f) + O(\log n)
\end{eqnarray*}
here, $\R_{\mathrm{priv}}(f)$ is also $\isr_0(f)$ and $\R(f)$ is also $\isr_1(f)$.
\end{theorem}

\subsection{Information Complexity}

Information complexity\footnote{we will often abbreviate ``information complexity'' by $\IC$} is an interactive analogue of {\em Shannon's information theory} \cite{shannon1948}. Informally, information complexity is defined as the minimum number of bits of {\em information} that the two parties have to reveal to each other, when the inputs $(\bx, \by) \in \bit^n \times \bit^n$ are coming from the `worst' possible distribution $\mu$.

\begin{defn}[(Prior-Free) Interactive Information Complexity; \cite{braverman2012interactive}]
For any $f : \bit^n \times \bit^n \to \bit$, the \emph{(prior-free) interactive information complexity} of $f$, denoted by $\IC(f)$, is defined as, 
$$\IC(f) \quad = \quad \inf_{\Pi} \ \ \sup_{\mu} \ \ I(X; \Pi|Y) + I(Y; \Pi|X)$$

\noindent where, the infimum is over all randomized protocols $\Pi$ such that for all $\bx, \by \in \bit^n$ such that $f(\bx, \by) \ne \ ?$, $\Pr[\Pi(\bx,\by) \ne f(\bx,\by)] \le 1/3$ and the supremum is over all distributions $\mu(\bx, \by)$ over $\bit^n \times \bit^n$. [$I(A;B|C)$ is the mutual information between $A$ and $B$ conditioned on $C$]
\end{defn}

We refer the reader to the survey by Weinstein \cite{weinstein_ICsurvey} for a more detailed understanding of the definitions and the role of information complexity in communication complexity.

A general question of interest is: what is the relationship between $\IC(f)$ and $\R(f)$? It is straightforward
to show $\R(f) \geq \IC(f)$. Upper bounding $\R(f)$ as a function of $\IC(f)$ has been 
investigated in several works including the work of Barak et al.~\cite{barak2013compress}.
The cleanest relation known is that $\R(f) \le 2^{O(\IC(f))}$ \cite{braverman2012interactive}. 
Our first result, namely Theorem~\ref{thm:main-1}, shows that for permutation-invariant functions, $\R(f)$ is {\em not much larger} than $\IC(f)$.

\fi

\ifnum \focs=0
\subsection{Some Useful Communication Problems}
\fi

\ifnum\focs=1
The basic problems we work with are parameterized variants of Gap Hamming Distance and Disjointness. We define them briefly
below.
For integers $a,b,c,g$ and $n$ the Gap Hamming Distance problem is given by $\GHD^n_{a,b,c,g}(\bx,\by) = 1$ if $|\bx|=a$, $|\by| = b$ and
$\GHD^n_{a,b,c,g}(\bx,\by) = 0$ if $|\bx|=a$, $|\by| = b$ and
$\Delta(x,y) \leq c-g$ and $?$ otherwise. $\eGHD$ is defined similarly with the inequalities on
$\Delta(x,y)$ replaced by equalities.
The Unique Disjointness problem is defined by 
$\UDISJ^n_t(\bx,\by) = 1$ if $|\bx|=|\by|=t$ and $|\bx\wedge\by| = 1$, 
$\UDISJ^n_t(\bx,\by) = 0$ if $|\bx|=|\by|=t$ and $|\bx\wedge\by| = 0$ and $?$ otherwise.
Finally we mention the Lopsided Unique Disjointness problem by 
$\LUDISJ^n_t(\bx,\by) = 1$ if $|\bx|=t$, $|\by|=1$ and $|\bx\wedge\by| = 1$, while
$\LUDISJ^n_t(\bx,\by) = 0$ if $|\bx|=t$, $|\by|=1$ and $|\bx\wedge\by| = 0$
and $?$ otherwise.

We use the following bounds on the complexity of the above problems.

\begin{proposition}
The following lower bounds hold:
\begin{enumerate}
\item 
For all $t, w \in \mathbb{N}$, $\IC(\UDISJ^{2t+w}_t) = \Omega(\min\set{t,w})$.
\item
For all $t, w \in \mathbb{N}$, $\R^{\oneway}(\LUDISJ^{t+w}_t) = \Omega(\min\set{t,w})$.
\end{enumerate}
\end{proposition}

These statements follow from the literature
~\cite{razborov1992distributional,kerenidis2012lower,jain2010partition,jayram2008one}. We include a detailed proof in the
full version of this paper~\cite{GKS:full}.

\else

Central to our proof techniques is a multi-parameter version of {\sc Gap-Hamming-Distance}, which we define as follows.

\begin{defn}[{\sc Gap-Hamming-Distance}, $\GHD^n_{a,b,c,g}$, $\eGHD^n_{a,b,c,g}$] \label{defn:GHD}
We define $\GHD^n_{a,b,c,g}$ as the following partial function,
$$\GHD^n_{a,b,c,g}(\bx, \by) = \infork{1 & \text{if } |\bx| = a, |\by| = b \text{ and } \Delta(\bx,\by) \ge c+g\\ 0 & \text{if } |\bx| = a, |\by| = b \text{ and } \Delta(\bx,\by) \le c-g\\ ? & \text{otherwise} }$$

\noindent Additionally we define $\eGHD^n_{a,b,c,g}$ as the following partial function,
$$\eGHD^n_{a,b,c,g}(\bx, \by) = \infork{1 & \text{if } |\bx| = a, |\by| = b \text{ and } \Delta(\bx,\by) = c+g\\ 0 & \text{if } |\bx| = a, |\by| = b \text{ and } \Delta(\bx,\by) = c-g\\ ? & \text{otherwise} }$$
\end{defn}

\noindent Informally, computing $\GHD^n_{a,b,c,g}$ is equivalent to the following problem:
\begin{itemize}
\item Alice is given $\bx \in \bit^n$ such that $|\bx| = a$
\item Bob is given $\by \in \bit^n$ such that $|\by| = b$
\item They wish to distinguish between the cases $\Delta(\bx, \by) \ge c+g$ and $\Delta(\bx, \by) \le c-g$.
\end{itemize}

In the $\eGHD^n_{a,b,c,g}$ problem, they wish to distinguish between the cases $\Delta(\bx, \by) = c+g$ or $\Delta(\bx, \by) = c-g$. Arguably, $\eGHD^n_{a,b,c,g}$ is an `easier' problem than $\GHD^n_{a,b,c,g}$. However, in this work we will show that in fact $\GHD^n_{a,b,c,g}$ is not much harder than $\eGHD^n_{a,b,c,g}$.\\

We will use certain known lower bounds on the information complexity and one-way communication complexity of $\eGHD^n_{a,b,c,g}$ for some settings of the parameters. The two main settings of parameters that we will be using correspond to the problems of {\sc Unique-Disjointness} and {\sc Sparse-Indexing} (a variant of the more well-known {\sc Indexing} problem).

\subsubsection{IC lower bound for {\sc Unique-Disjointness}} \label{subsec:udisj_lb}

\begin{defn}[{\sc {\sc Unique-Disjointness}}, $\UDISJ^n_t$] \label{defn:UDISJ}
$\UDISJ^n_t$ is given by the following partial function,
$$\UDISJ^n_t(\bx, \by) = \infork{1 & \text{if } |\bx| = t, |\by| = t \text{ and } |\bx \wedge \by| = 1\\ 0 & \text{if } |\bx| = t, |\by| = t \text{ and } |\bx \wedge \by| = 0\\ ? & \text{otherwise} }$$
\noindent Note that $\UDISJ^n_t$ is an instance of $\eGHD^n_{t,t,2t-1,1}$.
\end{defn}

Informally, $\UDISJ^n_t$ is the problem where the inputs $\bx, \by \in \set{0,1}^{n}$ satisfy $|\bx| = |\by| = t$ and Alice and Bob wish to decide whether $|\bx \wedge \by| = 1$ or $|\bx \wedge \by| = 0$ (promised that it is one of them is the case).

\begin{lem}\label{lem:basic_disj_lb}
For all $n \in \bbN$, $\IC(\UDISJ^{3n}_n) = \Omega(n)$.
\end{lem}

\begin{proof}
Bar-Yossef et al. \cite{BarYossefJKS04} proved that {\sc General-Unique-Disjointness}, that is, unique disjointness without restrictions on $|\bx|$, $|\by|$ on inputs of length $n$, has information complexity $\Omega(n)$. We convert the general {\sc Unique-Disjointness} instance into an instance of $\UDISJ^{3n}_n$ by a simple padding argument as follows. Given an instance of general {\sc Unique-Disjointness} $(\bx', \by') \in \bit^n \times \bit^n$. Alice constructs $\bx = \bx' \circ 1^{(n-|\bx'|)} \circ 0^{(n+|\bx'|)}$ and Bob constructs $\by = \by' \circ 0^{(n+|\by'|)} \circ 1^{(n-|\by'|)}$. Note that $|\bx \wedge \by| = |\bx' \wedge \by'|$. Also, we have that $\bx, \by \in \bit^{3n}$, and $|\bx| = |\by| = n$. Thus, we have reduced {\sc General-Unique-Disjointness} to $\UDISJ^{3n}_n$, and thus the lower bound of \cite{BarYossefJKS04} implies that $\IC(\UDISJ^{3n}_n) = \Omega(n)$.
\end{proof}

On top of the above lemma, we apply a simple {\em padding} argument in addition to the above lower bound, to get a more general lower bound for {\sc Unique-Disjointness} as follows.

\begin{proposition}[Unique Disjointness IC Lower Bound]\label{prop:u_disj_low_bd}
For all $t, w \in \mathbb{N}$, $$\IC(\UDISJ^{2t+w}_t) = \Omega(\min\set{t,w})$$
\end{proposition}
\begin{proof} We look at two cases, namely $w \le t$ and $w > t$.\\
{\bf Case 1.} [$w \le t$]: We have from Lemma~\ref{lem:basic_disj_lb} that $\IC(\UDISJ^{3w}_{w}) \ge \Omega(w)$. We map the instance $(\bx', \by')$ of $\UDISJ^{3w}_{w}$ to an instance $(\bx, \by)$ of $\UDISJ^{2t+w}_t$ by the following reduction, $\bx = \bx' \circ 1^{(t-w)} \circ 0^{(t-w)}$ and $\by = \by' \circ 0^{(t-w)} \circ 1^{(t-w)}$. This implies that $\IC(\UDISJ^{2t+w}_t) = \Omega(w)$.\\

\noindent {\bf Case 2.} [$w > t$]: We have from Lemma~\ref{lem:basic_disj_lb} that $\IC(\UDISJ^{3t}_{t}) = \Omega(t)$. As before, we map the instance $(\bx', \by')$ of $\UDISJ^{3t}_{t}$ to an instance $(\bx, \by)$ of $\UDISJ^{2t+w}_t$ by the following reduction, $\bx = \bx' \circ 0^{(w-t)}$ and $\by = \by' \circ 0^{(w-t)}$. This implies that $\IC(\UDISJ^{2t+w}_t) = \Omega(t)$.\\

\noindent Combining the above two lower bounds, we get that $\IC(\UDISJ^{2t+w}_t) = \Omega(\min\set{t,w})$.
\end{proof}

\subsubsection{1-way CC lower bound for {\sc Sparse-Indexing}}

\begin{defn}[{\sc Sparse-Indexing}, $\SI^{n}_t$] \label{defn:sparse_indexing}
$\SI^n_t$ is given by the following partial function,
$$\SI^n_t(\bx, \by) = \infork{1 & \text{if } |\bx| = t, |\by| = 1 \text{ and } |\bx \wedge \by| = 1\\ 0 & \text{if } |\bx| = t, |\by| = 1 \text{ and } |\bx \wedge \by| = 0\\ ? & \text{otherwise} }$$

\noindent Note that $\SI^n_t$ is an instance of $\eGHD^n_{t,1,t,1}$.
\end{defn}

Informally, $\SI^n_t$ is the problem where the inputs $\bx, \by \in \bit^{n}$ satisfy $|\bx| = t$ and $|\by| = 1$ and Alice and Bob wish to decide whether $|\bx \wedge \by| = 1$ or $|\bx \wedge \by| = 0$ (promised that one of them is the case).

\begin{lem}\label{lem:1way_basic_ludisj_lb}
For all $a \in \mathbb{N}$, $\R^{\oneway}(\SI^{2n}_n) = \Omega(n)$.
\end{lem}

\begin{proof}
Jayram et al. \cite{jayram2008one} proved that if Alice is given $\bx \in \{0,1\}^n$, Bob is given $i \in [n]$, and Bob needs to determine $x_i$ upon receving a single message from Alice, then Alice's message should consist of $\Omega(n)$ bits, even if they are allowed shared randomness. Using their result, we deduce that  $\R^{\oneway}(\SI^{2n}_n) = \Omega(n)$ via the following simple padding argument: Alice and Bob double the length of their strings from $n$ to $2n$, with Alice's new input consisting of $(\bx,\overline{\bx})$ while Bob's new input consists of $(\be_i, 0)$, where $\overline{\bx}$ is the bitwise complement of $\bx$ and $\be_i$ is the indicator vector for location $i$. Note that the Hamming weight of Alice's new string is equal to $n$ while its length is $2n$, as desired.
\end{proof}

On top of the above lemma, we apply a simple {\em padding} argument in addition to the above lower bound, to get a more general lower bound for {\sc Sparse-Indexing} as follows.

\begin{proposition}[{\sc Sparse-Indexing} 1-way CC Lower Bound]\label{prop:ludisj_low_bd}
For all $t, w \in \mathbb{N}$, $$\R^{\oneway}(\SI^{t+w}_t) = \Omega(\min\set{t,w})$$
\end{proposition}

\begin{proof} We look at two cases, namely $w \le t$ and $w > t$.\\
{\bf Case 1.} [$w \le t$]: We have from Lemma~\ref{lem:1way_basic_ludisj_lb} that $\R^{\oneway}(\SI^{2w}_{w}) \ge \Omega(w)$. We map the instance $(\bx', \by')$ of $\SI^{2w}_{w}$ to an instance $(\bx, \by)$ of $\SI^{t+w}_t$ by the following reduction, $\bx = \bx' \circ 1^{(t-w)}$ and $\by = \by' \circ 0^{(t-w)}$. This implies that $\R^{\oneway}(\SI^{t+w}_t) \ge \Omega(w)$.\\

\noindent {\bf Case 2.} [$w > t$]: We have from Lemma~\ref{lem:1way_basic_ludisj_lb} that $\R^{\oneway}(\SI^{2t}_{t}) = \Omega(t)$. We map the instance $(\bx', \by')$ of $\SI^{2t}_t$ to an instance $(\bx, \by)$ of $\SI^{t+w}_t$ by the following reduction, $\bx = \bx' \circ 0^{(w-t)}$ and $\by = \by' \circ 0^{(w-t)}$. This implies that $\R^{\oneway}(\SI^{t+w}_t) \ge \Omega(t)$.\\

\noindent Combining the above two lower bounds, we get that $\R^{\oneway}(\SI^{t+w}_t) \ge \Omega(\min\set{t,w})$.
\end{proof}



\section{Coarse Characterization of Information Complexity}\label{sec:ic_vs_cc}

In this section, we prove the first of our results, namely Theorem \ref{thm:main-1}, which we restate below for convenience of the reader.

\mainthmone*

\noindent where $m(f)$ is the combinatorial measure we define in Definition~\ref{def:measure_2way_partial}.

\subsection{Overview of proof}
\noindent We construct a measure $m(f)$ such that $m(f) \le \IC(f) \le \R(f) \le \Otilde(m(f)^4) + O(\log n)$. In order to do this, we look at the {\em slices} of $f$ obtained by restricting $|\bx|$ and $|\by|$. As in Observation~\ref{obs:interchang_rep}, let $h_{a,b}(\Delta(\bx, \by))$ be the restriction of $h$ to $|\bx| = a$ and $|\by| = b$. We define the notion of a {\em jump} in $h_{a,b}$ as follows.

\begin{defn}[Jump in $h_{a,b}$]\label{defn:jumps}
$(c,g)$ is a {\em jump} in $h_{a,b}$ if $h_{a,b}(c+g) \ne h_{a,b}(c-g)$, both $h_{a,b}(c+g)$, $h_{a,b}(c-g)$ are in $\bit$ and $h_{a,b}(r)$ is undefined for $c-g < r < c+g$.
\end{defn}

Thus, any protocol that computes $f$ with low error will in particular be able to solve a {\sc Gap-Hamming-Distance} problem $\eGHD^n_{a,b,c,g}$
\ifnum\focs=0
as in Definition~\ref{defn:GHD}.
\else
defined in Section~\ref{sec:prelim}.
\fi
Thus, if $(c,g)$ is a {\em jump} for $h_{a,b}$, then $\IC(\eGHD^n_{a,b,c,g})$ is a lower bound on $\IC(f)$. We will prove lower bounds on $\IC(\eGHD^n_{a,b,c,g})$ for any value of $a$, $b$, $c$ and $g$ by obtaining a variety of reductions from {\sc Unique-Disjointness}, and then our measure $m(f)$ will be obtained by taking the largest of these lower bounds for $\IC(\eGHD^n_{a,b,c,g})$ over all choices of $a$ and $b$ and jumps $(c,g)$ in $h_{a,b}$.\\

\noindent Suppose $m(f) = k$. We construct a randomized communication protocol with cost $\Otilde(k^4) + O(\log n)$ that computes $f$ correctly with low constant error. The protocol works as follows: First, Alice and Bob exchange the values $|\bx|$ and $|\by|$, which requires $O(\log n)$ communication (say, $|\bx| = a$ and $|\by| = b$). Now, all they need to figure out is the range in which $\Delta(\bx, \by)$ lies (note that finding $\Delta(\bx, \by)$ exactly can require $\Omega(n)$ communication!). Let $\calJ(h_{a,b}) = \set{(c_1, g_1), (c_2, g_2), \cdots, (c_m, g_m)}$ be the set of all {\em jumps} in $h_{a,b}$. Note that the intervals $[c_i - g_i, c_i + g_i]$ are all pairwise disjoint. To compute $h_{a,b}(\Delta(\bx, \by))$, it suffices for Alice and Bob need to {\em resolve} each jump, that is, for each $i \in [m]$, they need to figure out whether $\Delta(\bx, \by) \ge c+g$ or $\Delta(\bx, \by) \le c-g$. We will show that any particular jump can be resolved with a constant probability of error using $O(k^3)$ communication, and the number of jumps $m$ is at most $2^{O(k)} \log n$. Although the number of jumps is large, it suffices for Alice and Bob to do a binary search through the jumps, which will require them to resolve only $O(k \log \log n)$ jumps each requiring $\Otilde(k^3)$ communication. Thus, the total communication cost will be $\Otilde(k^4) + O(\log n)$.\footnote{We will need to resolve each jump correctly with error probability of at most $1/\Omega(k \log \log n)$. And for that we will actually require $O(k^3 \log k \log \log \log n)$ communication. So the total communication is really $O(k^4 \cdot \log k \cdot \log \log n \cdot \log \log \log n)$ which we write as $\Otilde(k^4) + O(\log n)$ in short. See Remark~\ref{remark:otilde_m}.}

\subsection{Proof of Theorem \ref{thm:main-1}}

As outlined earlier, we define the measure $m(f)$ as follows.

\begin{defn}[Measure $m(f)$]\label{def:measure_2way_partial}
Given a permutation-invariant function $f : \bits^n \times \bits^n \to \{0,1,?\}$ and integers $a$, $b$, s.t. $0\leq a,b \leq n$, let $h_{a,b}:\set{0,\cdots,n} \to \set{0,1,?}$ be the function given by $h_{a,b}(d) = f(\bx,\by)$ if there exist $\bx,\by$ with $|\bx|= a$, $|\by|=b$, $\Delta(\bx,\by) = d$ and $?$ otherwise. (Note. by permutation invariance of $f$, $h_{a,b}$ is well-defined.) Let $\calJ(h_{a,b})$ be the set of {\em jumps} in $h_{a,b}$, defined as follows,
$$\calJ(h_{a,b}) \quad \defeq \quad \setdef{(c,g)}{\inmat{h_{a,b}(c-g), \ h_{a,b}(c+g) \in \bit \\ h_{a,b}(c-g) \ne h_{a,b}(c+g) \\ \forall i \in (c-g, c+g) \ :\ h_{a,b}(i) = \ ?}}$$
Then, we define $m(f)$ as follows.
$$m(f) \quad \defeq \quad \frac{1}{C} \cdot \max_{\substack{a, b \in [n] \\ (c,g) \in \calJ(h_{a,b})}} \max\set{\frac{\min\set{a,b,c,n-a,n-b,n-c}}{g}, \log\inparen{\frac{\min \set{c, n-c}}{g}}}$$
where $C$ is a suitably large constant (which does not depend on $n$).
\end{defn}

\noindent We will need the following lemma to show that the measure $m(f)$ is a lower bound on $\IC(f)$.

\begin{lem}\label{lem:main_low_bds_ghd}
For all $n, a, b, c, g$ for which $\eGHD^n_{a,b,c,g}$ is a meaningful problem\footnote{that is, $a, b \le n$ and $c+g$ and $c-g$ are achievable hamming distances when $|\bx| = a$ and $|\by| = b$.}, the following lower bounds hold,
\begin{eqnarray*}
\IC(\eGHD^n_{a,b,c,g}) &\ge& \frac{1}{C} \cdot \frac{\min\inbrace{a,b,c,n-a,n-b,n-c}}{g}\\
\IC(\eGHD^n_{a,b,c,g}) &\ge& \frac{1}{C} \cdot \log\inparen{\frac{\min \set{c, n-c}}{g}}
\end{eqnarray*}
\end{lem}

\noindent Next, we obtain randomized communication protocols to solve $\GHD^n_{a,b,c,g}$.

\begin{lem}\label{lem:GHD_upper_bounds}
Let $a \le b \le n/2$. Then, the following upper bounds hold
\begin{eqnarray*}
\R(\GHD^n_{a,b,c,g}) &=& O\inparen{\inparen{\frac{a}{g}}^2\inparen{\log\inparen{\frac{b}{a}}+\log\inparen{\frac{a}{g}}}}\\
\R(\GHD^n_{a,b,c,g}) &=& O\inparen{\min\set{\inparen{\frac{c}{g}}^2, \inparen{\frac{n-c}{g}}^2}}
\end{eqnarray*}
\end{lem}

\noindent We defer the proofs of the Lemmas~\ref{lem:main_low_bds_ghd} and \ref{lem:GHD_upper_bounds} to Section \ref{sec:GHD_results}. For now, we will use these lemmas to prove Theorem~\ref{thm:main-1}. First, we show that each {\em jump} can be resolved using $O(m(f)^3)$ communication.

\begin{lem}\label{lem:resolve_jump}
Suppose $m(f) = k$, and let $|\bx| = a$ and $|\by| = b$. Let $(c,g)$ be any jump in $h_{a,b}$. Then the problem of $\GHD^n_{a,b,c,g}$, that is, deciding whether $\Delta(\bx, \by) \ge c+g$ or $\Delta(\bx,\by) \le c-g$ can be solved, with a constant probability of error, using $O(k^3)$ communication.
\end{lem}

\ifnum\focs=0
\begin{proof}
We can assume without loss of generality that $a \le b \le n/2$. This is because both Alice and Bob can flip their respective inputs to ensure $a, b \le n/2$. Also, if $a > b$, then we can flip the role of Alice and Bob to get $a \le b \le n/2$.

\noindent Since $a \le b \le n/2$, from the definition of $m(f)$ in Lemma~\ref{lem:main_low_bds_ghd} we have that,
$$\min\set{\frac{a}{g}, \frac{c}{g}, \frac{n-c}{g}} \le O(k)$$

\noindent We consider two cases as follows.

\noindent {\bf Case 1.} $(c/g) \le O(k)$ or $((n-c)/g) \le O(k)$: In this case we have, from part (ii) of Lemma~\ref{lem:GHD_upper_bounds}, a randomized protocol with cost $O\inparen{\min\set{(c/g)^2, ((n-c)/g)^2}} = O(k^2)$.

\noindent {\bf Case 2.} $(a/g) \le O(k)$: In this case, we will show that $\inparen{(a/g)^2\inparen{\log\inparen{b/a}+\log\inparen{a/g}}} \le O(k^3)$. Clearly, $(a/g)^2 \le O(k^2)$ and $\log(a/g) \le O(\log k)$. We will now show that in fact $\log(b/a) \le O(k)$. From part (ii) of Lemma~\ref{lem:main_low_bds_ghd} we know that either $\log(c/g) \le O(k)$ or $\log((n-c)/g) \le O(k)$. Thus, it suffices to show that $(b/a) \le O\inparen{\min\set{(c/g), ((n-c)/g)}}$. We know that $b - a + g \le c \le b + a - g$. The left inequality gives $(b/a) \le (c + a - g)/a = (c/g).(g/a) + 1 - (g/a) \le O(c/g)$ (since $g/a \le 1$). The right inequality gives $((n-b)/a) \le ((n-c)/g).(g/a) + 1 - g/a \le O((n-c)/g)$. Since $b \le n/2$, we have $(b/a) \le ((n-b)/a) \le O((n-c)/g)$.
\end{proof}
\fi

\noindent Next, we obtain an upper bound on $|\calJ(h_{a,b})|$, that is, the number of jumps in $h_{a,b}$.

\begin{lem}\label{lem:upper_bds_jumps}
For any function $f : \set{0,1}^n \times \set{0,1}^n \rightarrow \set{0,1}$ with $m(f) = k$, the number of {\em jumps} in $h_{a,b}$ is at most $2^{O(k)}\log n$.
\end{lem}

\ifnum\focs=1
\noindent The proofs of Lemmas~\ref{lem:main_low_bds_ghd}, \ref{lem:GHD_upper_bounds}, \ref{lem:resolve_jump} and \ref{lem:upper_bds_jumps} are fairly involved, and they appear in the full version along with the full details.\\ 

\noindent {\bf Remark.} It is a valid concern that we are hiding a $\log \log n$ factor in the $\Otilde(m(f)^4)$ term. But this is fine because of the following reason: If $m(f) \le (\log n)^{1/5}$, then the $O(\log n)$ term is the dominating term and thus the overall communication is $O(\log n)$. And if $m(f) \ge (\log n)^{1/5}$, then $\log m(f) = \Omega(\log \log n)$, in which case the $\Otilde(m(f)^4)$ is truly hiding only $\poly \log m(f)$ factors.
\fi

\ifnum\focs=0
\begin{proof}
Let $\calJ = \set{(c_1, g_1), \cdots, (c_m, g_m)}$ be the set of all jumps in $h_{a,b}$. Partition $\calJ$ into $\calJ_1~\union~\calJ_2$, where $\calJ_1 = \setdef{(c,g) \in \calJ}{c \le n/2}$ and $\calJ_2 = \setdef{(c,g) \in \calJ}{c > n/2}$. From second part in Lemma~\ref{lem:main_low_bds_ghd} we know the following:
\begin{eqnarray*}
\forall (c,g) \in \calJ_1 \quad &:& \quad \log \inparen{\frac{c}{g}} \le O(k) \quad \text{that is,} \quad g \ge c2^{-O(k)}\\
\forall (c,g) \in \calJ_2 \quad &:& \quad \log \inparen{\frac{n-c}{g}} \le O(k) \quad \text{that is,} \quad g \ge (n-c) 2^{-O(k)}
\end{eqnarray*}
Let $\calJ_1 = \set{(c_1, g_1), \cdots, (c_p, g_p)}$, where the $c_i$'s are sorted in increasing order. We have that $c_i + g_i \le c_{i+1} - g_{i+1}$ for all $i$. That is, we have that $c_i(1 + 2^{-O(k)}) \le c_{i+1}(1-2^{-O(k)})$, which gives that $c_{i+1} \ge c_i (1 + 2^{-O(k)})$. Thus, $n/2 \ge c_p \ge c_1(1+2^{-O(k)})^p$, which gives that $|\calJ_1| = p \le 2^{O(k)} \log n$. Similarly, by looking at $n-c_i$'s in $\calJ_2$, we get that $|\calJ_2| \le 2^{O(k)} \log n$, and thus, $|\calJ| \le 2^{O(k)} \log n$.
\end{proof}

\noindent We now complete the proof of Theorem~\ref{thm:main-1}. 

\begin{proofof}{Theorem~\ref{thm:main-1}}
Any protocol to compute $f$ also computes $\eGHD^n_{a,b,c,g}$ for any $a$, $b$ and any jump $(c,g) \in \calJ(h_{a,b})$. Consider the choice of $a$, $b$ and a jump $(c,g) \in \calJ(h_{a,b})$ such that the lower bound obtained on $\IC(\eGHD^n_{a,b,c,g})$ through Lemma~\ref{lem:main_low_bds_ghd} is maximized, which by definition is equal to $m(f)$ (after choosing an appropriate $C$ in the definition of $m(f)$). Thus, we have $m(f) \le \IC(\eGHD^n_{a,b,c,g}) \le \IC(f)$.

\noindent We also have a protocol to solve $f$, which works as follows: First Alice and Bob exchange $|\bx| = a$ and $|\by| = b$, requiring $O(\log n)$ communication. From Lemma~\ref{lem:upper_bds_jumps}, we know that the number of jumps in $h_{a,b}$ is at most $2^{O(m(f))} \log n$, and so Alice and Bob need to do a binary search through the jumps, resolving only $O(m(f) \log \log n)$ jumps, each with an error probability of at most $O(1/m(f) \log \log n)$. This can be done using $\Otilde(m(f)^3)$ communication\footnote{Here, $\Otilde(m(f)^3) = O(m(f)^3 \log m(f) \log \log \log n)$.} (using Lemma~\ref{lem:resolve_jump}). Thus, the total amount of communication is $\R(f) \le \Otilde(m(f)^4) + O(\log n)$\footnote{Here $\Otilde(m(f)^4) = O(m(f)^4 \log m(f) \log \log n \log \log \log n$)}. All together, we have shown that,
$$m(f) \le \IC(f) \le \R(f) \le \Otilde(m(f)^4) + O(\log n)$$
\end{proofof}


\begin{remark} \label{remark:otilde_m}
It is a valid concern that we are hiding a $\log \log n$ factor in the $\Otilde(m(f)^4)$ term. But this is fine because of the following reason: If $m(f) \le (\log n)^{1/5}$, then the $O(\log n)$ term is the dominating term and thus the overall communication is $O(\log n)$. And if $m(f) \ge (\log n)^{1/5}$, then $\log m(f) = \Omega(\log \log n)$, in which case the $\Otilde(m(f)^4)$ is truly hiding only $\poly \log m(f)$ factors.
\end{remark}

\noindent In the following section, we prove the main technical lemmas used, namely Lemmas~\ref{lem:main_low_bds_ghd} and \ref{lem:GHD_upper_bounds}.

\subsection{Lower and upper bounds on Gap Hamming Distance} \label{sec:GHD_results}

In this section, we prove lower bounds on information complexity (Lemma~\ref{lem:main_low_bds_ghd}), and upper bounds on the randomized communication complexity (Lemma~\ref{lem:GHD_upper_bounds}) of {\sc Gap-Hamming-Distance}.

\subsubsection*{Lower bounds}
\noindent We will prove Lemma~\ref{lem:main_low_bds_ghd} by getting certain reductions from {\sc Unique-Disjointness} (namely Proposition~\ref{prop:u_disj_low_bd}). In order to do so, we first prove lower bounds on information complexity of two problems, namely, {\sc Set-Inclusion} (Definition~\ref{defn:set_inclusion}) and {\sc Sparse-Indexing} (Definition~\ref{defn:sparse_indexing}). We do this by obtaining reductions from {\sc Unique-Disjointness}.

\begin{defn}[$\setinc^n_{p,q}$]\label{defn:set_inclusion}
Let $p \le q \le n$. Alice is given $\bx \in \set{0,1}^n$ such that $|\bx| = p$ and Bob is given $\by \in \set{0,1}^n$ such that $|\by| = q$ and they wish to distinguish between the cases $|\bx \wedge \by| = p$ and $|\bx \wedge \by| = p-1$. Note that $\setinc^n_{p,q}$ is the same as $\eGHD^n_{p,q,q-p+1,1}$.
\end{defn}

\begin{proposition}[{\sc Set-Inclusion} lower bound] \label{prop:set_inc_low_bd}
$\forall \, t, w \in \bbN$, $\IC(\setinc^{2t+w}_{t, t+w}) \ge \Omega(\min(t, w))$
\end{proposition}
\begin{proof}
We know that $\IC(\UDISJ^{2t+w}_t) \ge \Omega(\min(t,w))$ from Proposition \ref{prop:u_disj_low_bd}. Note that $\UDISJ^{2t+w}_t$ is same as the problem of $\eGHD^{2t+w}_{t,t,2t-1,1}$. If we instead think of Bob's input as complemented, we get that solving $\eGHD^{2t+w}_{t,t,2t-1,1}$ is equivalent to solving $\eGHD^{2t+w}_{t,t+w,w+1,1}$, which is same as $\setinc^{2t+w}_{t,t+w}$. Thus, we conclude that $\IC(\setinc^{n}_{t,t+w}) \ge \Omega(\min(t,w))$.
\end{proof}


\begin{proposition}[{\sc Sparse-Indexing} lower bound] \label{prop:sparse_index_low_bd}
$\forall \, t \in \bbN$, $\IC(\SI^{2^{t+1}}_{2^t}) \ge \Omega(t)$
\end{proposition}

\begin{proof}
We know that $\IC(\UDISJ^{t+1}_{t/3}) \ge \Omega(t)$ from Proposition \ref{prop:u_disj_low_bd}. Recall that $\SI^{2^{t+1}}_{2^t}$ is an instance of $\eGHD^{2^{t+1}}_{2^t,1,2^t,1}$. Alice uses $\bx$ to obtain the Hadamard code of $\bx$, which is $\bX \in \set{0,1}^{2^{t+1}}$ such that $\bX(\ba) = \ba\cdot\bx$ (for $\ba \in \bit^{t+1}$). On the other hand, Bob uses $\by$ to obtain the indicator vector of $\by$, which is $\bY \in \set{0,1}^{2^{t+1}}$ such that $\bY(\ba) = 1$ iff $\ba = \by$. Clearly $|\bY| = 1$. Observe that, $|\bX| = 2^{t}$ and $\bX(\by) = \by\cdot\bx$ is $1$ if $|\bx \wedge \by| = 1$ and $0$ if $|\bx \wedge \by| = 0$.  Thus, we end up with an instance of $\SI^{2^{t+1}}_{2^t}$. Hence $\IC(\SI^{2^{t+1}}_{2^t}) \ge \IC(\UDISJ^{t+1}_{t/3}) \ge \Omega(t)$.
\end{proof}

\noindent We now state and prove a technical lemma that will help us prove Lemma~\ref{lem:main_low_bds_ghd}.

\begin{lem}\label{le:first_low_bds_ghd}
Let $a \le b \le n/2$. Then, the following lower bounds hold,
\begin{enumerate}
\item[(i)] $\IC(\eGHD^n_{a,b,c,g}) \ge \Omega\inparen{\min\inbrace{\frac{c-b+a}{g}, \frac{n-c}{g}}}$
\item[(ii)] $\IC(\eGHD^n_{a,b,c,g}) \ge \Omega\inparen{\min\inbrace{\frac{a+b-c}{g}, \frac{c+b-a}{g}}}$
\item[(iii)] $\IC(\eGHD^n_{a,b,c,g}) \ge \Omega\inparen{\min\inbrace{\log \inparen{\frac{c}{g}}, \log \inparen{\frac{n-c}{g}}}}$
\end{enumerate}
\end{lem}

\begin{proof}
We prove the three parts of the above lemma using reductions from $\UDISJ$, $\setinc$ and $\SI$ respectively. Note that once we fix $|\bx| = a$ and $|\by| = b$, a jump $(c,g)$ is meaningful only if $b-a+g \le c \le b+a-g$ (since $b-a \le \Delta(\bx, \by) \le b+a$). We will assume that $c \equiv b+a-g (\mod 2)$, so that $c+g$ and $c-g$ will be achievable Hamming distances.\\

\noindent {\bf Proof of (i).} We obtain a reduction from $\UDISJ^{3t}_t$ for which we know from Proposition~\ref{prop:u_disj_low_bd} that $\IC(\UDISJ^{3t}_t)~\ge~\Omega(t)$. Recall that $\UDISJ^{3t}_t$ is same as $\eGHD^{3t}_{t,t,2t-1,1}$. Given any instance of $\eGHD^{3t}_{t,t,2t-1,1}$, we first repeat the instance $g$ times to get an instance of $\eGHD^{3gt}_{gt, gt, g(2t-1), g}$. Now, we need to append $(a-gt)$ 1's to $\bx$ and $(b-gt)$ 1's to $\by$. This will increase the Hamming distance by a fixed amount which is at least $(b-a)$ and at most $(b+a-2gt)$. Also, the number of inputs we need to add is at least $((a-gt)+(b-gt)+(c-g(2t-1)))/2$ \footnote{We will be repeatedly using this idea in several proofs. The reason we obtain the said constraints is as follows: Suppose Alice has to add $A$ $1$'s to her input and Bob has to add $B$ $1$'s to his input. Then the hamming distance increases by an amount $C$ such that $|A-B| \le C \le A+B$. Also, the minimum number of coordinates that need to be added to achieve this is at least $(A+B+C)/2$}. Thus, we can get a reduction to $\eGHD^n_{a,b,c,g}$ if and only if,
$$(b-a) \le c - g(2t-1) \le b+a-2gt$$
$$n \ge 3gt + \frac{(a-gt)+(b-gt)+(c-g(2t-1))}{2}$$

\noindent The constraints on $c$ give us that $2gt \le c - (b-a) + g$ and $c \le b+a-g$ (recall that this is always true). The constraint on $n$ gives that $gt \le n - (a+b+c+g)/2$, which is equivalent to $$t \le \frac{n-a-b}{2g} + \frac{n-c-g}{2g}$$
Thus, we can have the reduction work by choosing $t$ to be
$$t = \min\inbrace{\frac{c-b+a+g}{2g},\frac{n-c-g}{2g}} = \Omega\inparen{\min\inbrace{\frac{c-b+a}{g}, \frac{n-c}{g}}}$$
(since $n \ge a+b$) and thus we obtain $$\IC(\eGHD^n_{a,b,c,g}) \ge \IC(\UDISJ^{3t}_t) \ge \Omega\inparen{\min\inbrace{\frac{c-b+a}{g}, \frac{n-c}{g}}}$$

\noindent {\bf Proof of (ii).} We obtain a reduction from $\setinc^{m}_{t,w}$ (where $m = 2t+w$) for which we know from Proposition~\ref{prop:set_inc_low_bd} that $\IC(\setinc^{m}_{t,w})~\ge~\Omega(\min\set{t,w})$. Recall that $\setinc^{m}_{t,w}$ is same as $\eGHD^{m}_{t,t+w,w+1,1}$. Given an instance of $\eGHD^{m}_{t,t+w,w+1,1}$, we first repeat the instance $g$ times to get an instance of $\eGHD^{gm}_{gt,gt+gw,gw+g,g}$. Now, we need to append $(a-gt)$ 1's to $\bx$ and $(b-gt-gw)$ 1's to $\by$. This will increase the Hamming distance by a fixed amount which is at least $|b-a-gw|$ and at most $(b-gt-gw)+(a-gt)$. Also, the number of inputs we need to add is at least $((a-gt)+(b-gt-gw)+(c-g(w+1)))/2$. Thus, we can get a reduction to $\eGHD^n_{a,b,c,g}$ if and only if,
$$|b-a-gw| \le c -g(w+1) \le b+a-2gt-gw$$
$$n \ge 2gt+gw+ \frac{(b-gt-gw)+(a-gt)+(c-gw-g)}{2}$$

\noindent The left constraint on $c$ requires $c \ge \max\set{b-a+g, 2gw-(b-a)+g}$. We know that $c \ge b-a+g$, so the only real constraint is $c \ge 2gw - (b-a) + g$, which gives us that,
$$w \le \frac{c+b-a-g}{2g}$$
The right constraint on $c$ requires $c \le b+a-2gt+g$, which gives us that,
$$t \le \frac{a+b-c+g}{2g}$$
Suppose we choose $t = \frac{a+b-c+g}{2g}$. Then the constraint on $n$ is giving us that,
$$n \ge gt + \frac{a+b+c-g}{2} = \frac{a+b-c+g}{2} + \frac{a+b+c-g}{2} = a+b$$
We already assumed that $a \le b \le n/2$, and hence this is always true.

\noindent Thus, we choose $t = \frac{a+b-c+g}{2g}$ and $w = \frac{c+b-a-g}{2g}$, and invoking Proposition~\ref{prop:set_inc_low_bd}, we get, $$\IC(\eGHD^n_{a,b,c,g}) \ge \IC(\setinc^{2t+w}_{t,w}) \ge \min(\set{t,w}) \ge \Omega\inparen{\min\inbrace{\frac{a+b-c}{g}, \frac{c+b-a}{g}}}$$

\noindent {\bf Proof of (iii).} We obtain a reduction from $\SI^{2^{t+1}}_{2^t}$ for which we know from Proposition~\ref{prop:sparse_index_low_bd} that $\IC(\SI^{2^{t+1}}_{2^t})~\ge~\Omega(t)$. Recall that $\SI^{2^{t+1}}_{2^t}$ is same as $\eGHD^{2^{t+1}}_{2^t,1,2^t,1}$, which is equivalent to $\eGHD^{2^{t+1}}_{1,2^t,2^t,1}$ (if we flip roles of Alice and Bob). Given an instance of $\eGHD^{2^{t+1}}_{1,2^{t},2^{t},1}$, we first repeat the instance $g$ times to get an instance of $\eGHD^{g2^{t+1}}_{g,g2^{t},g2^{t},g}$. Now, we need to append $(a-g)$ 1's to $\bx$ and $(b-g2^{t})$ 1's to $\by$. This will increase the Hamming distance by a fixed amount which is at least $|b-g2^t-a+g|$ and at most $(b-g2^t+a-g)$. Also, the number of inputs we need to add is at least $((a-g)+(b-g2^t)+(c-g2^t))/2$. Thus, we can get a reduction to $\eGHD^n_{a,b,c,g}$ if and only if,

$$|b-g2^t-a+g| \le c - g2^t \le (b-g2^t+a-g)$$
$$n \ge g2^{t+1} + \frac{(a-g)+(b-g2^t)+(c-g2^t)}{2}$$

\noindent The left constraint on $c$ requires $c \ge \max\inbrace{b-a+g, 2g2^t - b + a - g}$. Since $c \ge b-a+g$ anyway, this only requires $2g \cdot 2^t \le c + b - a + g$. The right constraint on $c$ requires $c \le b+a-g$ which is also true anyway. The constraint on $n$ is equivalent to,
$$g2^t \le n - \frac{a+b+c-g}{2} = \frac{n-a-b}{2} + \frac{n-c-g}{2}$$

\noindent Thus, we choose $t$ such that,
$$t = \min\inbrace{\log_2 \inparen{\frac{c+b-a+g}{2g}}, \log_2 \inparen{\frac{n-c-g}{2g}}} \ge \Omega\inparen{\min{\frac{c}{g}, \frac{n-c}{g}}}$$
and invoking Proposition~\ref{prop:sparse_index_low_bd}, we get,
$$\IC(\eGHD^n_{a,b,c,g}) \ \ge \ \IC(\SI^{2^{t+1}}_{2^t}) \ \ge \ \Omega(t) \ \ge \ \Omega\inparen{\min\inbrace{\log_2 \inparen{\frac{c}{g}}, \log_2 \inparen{\frac{n-c}{g}}}}$$
\end{proof}

\noindent We are now finally able to prove Lemma~\ref{lem:main_low_bds_ghd}.

\begin{proofof}{Lemma~\ref{lem:main_low_bds_ghd}}
Assume for now that $a \le b \le n/2$. From parts (i) and (ii) of Lemma~\ref{le:first_low_bds_ghd}, we know the following,
\begin{eqnarray*}
\IC(\eGHD^n_{a,b,c,g}) &\ge& \Omega\inparen{\min\inbrace{\frac{c-b+a}{g}, \frac{n-c}{g}}}\\
\IC(\eGHD^n_{a,b,c,g}) &\ge& \Omega\inparen{\min\inbrace{\frac{a+b-c}{g}, \frac{c+b-a}{g}}}
\end{eqnarray*}

\noindent Adding these up, we get that,
$$\IC(\eGHD^n_{a,b,c,g}) \ge \Omega\inparen{\min\inbrace{\frac{c-b+a}{g}, \frac{n-c}{g}} + \min\inbrace{\frac{a+b-c}{g}, \frac{c+b-a}{g}}}$$

\noindent Since $\min\set{A,B} + \min\set{C,D} = \min\set{A+C, A+D, B+C, B+D}$, we get that,
$$\IC(\eGHD^n_{a,b,c,g}) \ge \Omega\inparen{\min\inbrace{\frac{2a}{g}, \frac{2c}{g}, \frac{n+a+b-2c}{g}, \frac{n+b-a}{g}}}$$
For the last two terms, note that, $n+a+b-2c \ge n-c$ (since $a+b \ge c$) and $n+b-a \ge n$ (since $b \ge a$). Thus, overall we get,
$$\IC(\eGHD^n_{a,b,c,g}) \ge \Omega\inparen{\min\inbrace{\frac{a}{g}, \frac{c}{g}, \frac{n-c}{g}}}$$

\noindent Note that this was assuming $a \le b \le n/2$. In general, we get,
$$\IC(\eGHD^n_{a,b,c,g}) \ge \Omega\inparen{\frac{\min\inbrace{a,b,c,n-a,n-b,n-c}}{g}}$$
[We get $b$, $(n-b)$, $(n-a)$ terms because we could have flipped inputs of either or both of Alice and Bob. Moreover, to get $a \le b$, we might have flipped the role of Alice and Bob.]\\

\noindent The second lower bound of $\IC(\eGHD^n_{a,b,c,g}) \ge \Omega\inparen{\log \inparen{\frac{\min\inbrace{c,\ n-c}}{g}}}$ follows immediately from part (iii) of Lemma~\ref{le:first_low_bds_ghd}.

\noindent We choose $C$ to be a large enough constant, so that the desired lower bounds hold.
\end{proofof}

\subsubsection*{Upper bounds}

\noindent We will now prove Lemma~\ref{lem:GHD_upper_bounds}.\\

\begin{proofof}{Lemma~\ref{lem:GHD_upper_bounds}} We use different protocols to prove the two parts of the lemma.

\noindent {\bf Proof of Part 1.} The main idea is similar to that of Proposition $5.7$ of \cite{CGMS_ISR}, except that we first hash into a small number of buckets. The details are as follows.

Alice and Bob have an instance $(\bx, \by)$ of $\GHD^n_{a,b,c,g}$, that is, Alice has $\bx \in \bit^n$ such that $|\bx| = a$, and Bob has $\by \in \bit^n$ such that $|\by| = b$, and they wish to distinguish between the cases $\Delta(\bx,\by) \geq c+g$ and $\Delta(\bx,\by) \le c-g$.

Bob defines $\wtilde{\by} \in \set{1, -1}^n$ such that $\wtilde{y}_i = 1-2y_i$ for every $i \in [n]$. Then, the number of $-1$ coordinates in $\wtilde{\by}$ is exactly equal to $b$. It is easy to see that $\inangle{\bx, \wtilde{\by}} = (\Delta(\bx, \by) - b)$, and hence computing $\GHD^n_{a,b,c,g}(\bx, \by)$ is equivalent to distinguishing between the cases $\inangle{\bx , \wtilde{\by}} \geq \alpha a$ and $\inangle{\bx , \wtilde{\by}} \le \beta a$, where $\alpha \defeq (c-b+g)/a$ and $\beta \defeq (c-b-g)/a$. Note that $\alpha, \beta \in [-1,+1]$.

Alice and Bob use their shared randomness to generate a uniformly random hash function $h : [n] \to [B]$ where $B \defeq 100(a+b)(a/g)^2$. Basically, each coordinate $i \in [n]$ is mapped to one of the $B$ `buckets' uniformly and independently at random. Let $\supp(\bx):= \setdef{ i \in [n]}{x_i = 1}$. We say that a coordinate $i \in \supp(\bx)$ is {\em bad} if there is a coordinate $j \in [n]$, $j \neq i$ such that $h(j) = h(i)$ and at least one of $x_j = 1$ or $y_j =1$. For any $i \in \supp(\bx)$, the probability that $i$ is {\em bad} is at most $\inparen{1 - (1-1/B)^{(a+b)}} \le (a+b)/B = g^2/100a^2$. Thus, the expected number of {\em bad} coordinates is at most $(g^2/100a)$, and hence by Markov's inequality, we have that with probability at least $1-g/(10a)$, there are at least $a(1-g/(10a))$ coordinates in $\supp(x)$ that are not {\em bad}. Suppose we have chosen an $h$ such that at least $a(1-g/(10a))$ coordinates in $\supp(\bx)$ that are not {\em bad}. Let $\ell \defeq (2B/a)\ln(20a/g)$ and consider the following atomic protocol $\tau_h(\bx, \by)$:

\begin{itemize}
\item Alice and Bob use shared randomness to sample a sequence of $\ell$ indices $b_1, \dots, b_{\ell} \in_R [B]$.
\item Alice picks the smallest index $j \in [\ell]$ such that $h^{-1}(b_{j}) \cap \supp(\bx)$ is non-empty and sends $j$ to Bob. If there is no such $j$, then the protocol aborts and outputs $+1$.
\item Bob outputs $\prod\limits_{i \in h^{-1}(b_j)} \wtilde{y}_i$.
\end{itemize}
We first show that the difference in the probability of $\tau_h(\bx, \by)$ outputting $+1$, in the two cases $\inangle{\bx, \wtilde{\by}} \ge \alpha a$ and $\inangle{\bx, \wtilde{\by}} \le \beta a$, is at least $\Omega(g/a)$. In particular, we will show the following,
\begin{eqnarray}
\Pr[\tau_h(\bx, \by) = +1 \ | \ \inangle{\bx , \wtilde{\by}} \ge \alpha a] &\ge& 1/2+\alpha/2-3g/20a \label{eqn:alpha}\\
\Pr[\tau_h(\bx, \by) = +1 \ | \ \inangle{\bx , \wtilde{\by}} \le \beta a] &\le& 1/2+\beta/2+3g/20a \label{eqn:beta}
\end{eqnarray}
Before we prove Inequalities~\ref{eqn:alpha} and \ref{eqn:beta}, we first show how to use these to obtain our desired protocol. We observe that the difference in the two probabilities is at least $(\alpha - \beta)/2 - 3g/10a \ge \Omega(g/a)$. We repeat the above atomic procedure $T$ times and declare the input to satisfy $\inangle{\bx, \wtilde{\by}} \ge \alpha a$, if the number of times the output is $+1$, is at least $((1 + \alpha + \beta)/2)T$, and $\inangle{\bx, \wtilde{\by}} \le \beta a$ otherwise. A Chernoff bound implies that we will have the correct value the given $\GHD^n_{a,b,c,g}$ instance with probability at least $1 - e^{-\Omega(g^2T/a^2)}$. Setting $T = \Theta((a/g)^2)$ gives us that our protocol gives the right answer with probability at least $3/4$. And the overall communication complexity of this protocol is
$$O((a/g)^2 \log_2(\ell)) = O((a/g)^2 \log_2(2B/a \log_2(10a/g))) = O((a/g)^2(\log(b/a)+\log(a/g)))$$

\noindent We now prove Inequalities~\ref{eqn:alpha} and \ref{eqn:beta}. Note that there are three possible situations that can arise in $\tau_h(\bx, \by)$,
\begin{enumerate}
\item the protocol aborts (that is, for all $j \in [\ell]$, $h^{-1}(b_j) \cap \supp(\bx) = \emptyset$)
\item the index $j$ picked by Alice is such that $|h^{-1}(b_j) \cap \supp(\bx)| > 1$
\item the index $j$ picked by Alice is such that $|h^{-1}(b_j) \cap \supp(\bx)| = 1$
\end{enumerate}

\noindent We will say that an index $b \in [B]$ is `good' if $|h^{-1}(b) \cap \supp(\bx)| = 1$, `bad' if $|h^{-1}(b) \cap \supp(\bx)| > 1$, and `empty' if $|h^{-1}(b) \cap \supp(\bx)| = 0$

For Inequality~\ref{eqn:alpha}, we have that $\inangle{\bx, \wtilde{\by}} \ge \alpha a$, and we wish to lower bound the probability that the protocol $\tau_h(\bx, \by)$ outputs $+1$. Notice that, when the protocol aborts, it always outputs $+1$. And conditioned on not aborting, the probability that we are in situation (3) and not (2), is at least $(1 - g/10a)$. This is because the number of non-`empty' $b$'s is at most $a$, but the number of `good' $b$'s is at least $a(1-g/10a)$. Thus, overall, we get that,
\begin{eqnarray*}
\Pr[\tau_h(\bx, \by) = +1 \ | \ \inangle{\bx , \wtilde{\by}} \ge \alpha a] &\ge& \inparen{1 - \frac{g}{10a}}\inparen{\frac{1}{2} + \frac{\alpha}{2}}\\
&\ge& \inparen{\frac{1}{2} + \frac{\alpha}{2} - \frac{g}{10a}} \quad [\because \alpha \le 1]\\
&\ge& \inparen{\frac{1}{2} + \frac{\alpha}{2} - \frac{3g}{20a}}
\end{eqnarray*}

For Inequality~\ref{eqn:beta}, we have that $\inangle{\bx , \tilde{\by}} \le \beta a$, and we wish to upper bound the probability that the protocol $\tau_h(\bx, \by)$ outputs $+1$. The probability that we are in situation (1) is at most $\inparen{1 - \frac{(a-g/10)}{B}}^\ell \le \inparen{1 - \frac{a}{2B}}^\ell \le e^{-a\ell/2B} = g/20a$. Conditioned on not aborting, the probability that we are in situation (3) and not (2), is at least $(1-g/10a)$ as before. Thus overall, we get that,
\begin{eqnarray*}
\Pr[\tau_h(\bx, \by) = +1 \ | \ \inangle{\bx , \wtilde{\by}} \le \beta a] &\le& \inparen{\frac{g}{20a}} + \inparen{\frac{g}{10a}} + \inparen{1 - \frac{g}{10a}}\inparen{\frac{1}{2} + \frac{\beta}{2}}\\
&\le& \inparen{\frac{1}{2} + \frac{\beta}{2} + \frac{3g}{10a}}
\end{eqnarray*}

\noindent {\bf Proof of Part 2.} Kushilevitz, Ostrovsky and Rabani \cite{KOR00Sketching} gave a protocol for distinguishing between the cases $\Delta(\bx, \by) \ge c+g$ and $\Delta(\bx, \by) \le c-g$ using $O((c/g)^2)$ communication (without requiring any knowledge about $|\bx|$ and $|\by|$). The upper bound of $O(((n-c)/g)^2)$ follows by Alice flipping her input so that the task is of distinguishing between the cases $\Delta(\bx, \by) \ge n-c+g$ and $\Delta(\bx,\by) \le n-c-g$, and then the upper bound of \cite{KOR00Sketching} applies again.
\end{proofof}
\fi


\section{Communication with Imperfectly Shared Randomness}\label{sec:isr}

\ifnum\focs=0

In this section, we prove our second result, namely Theorem \ref{thm:main-2+3}, which we restate below for convenience of the reader.

\mainthmtwo*

The outline of this section is as follows: In Section~\ref{sec:atomic_isr_ub}, we prove upper bounds on the ISR-CC of two basic problems: {\sc Small-Set-Intersection} (in Section~\ref{subsec:ssi_isr}) and {\sc Small-Hamming-Distance} (in Section~\ref{subsec:k_HD_isr}). As an aside, in Section~\ref{subsec:strongly_perm_inv_subsec}, we introduce a new class of functions, called {\em strongly permutation-invariant functions}, which is a generalization of both {\sc Set-Intersection} and {\sc Hamming-Distance}, and show that for every strongly permutation-invariant function $f$ there is a polynomial relationship between $\R(f)$ and $\isr(f)$ (Theorem~\ref{thm:strong_pi}).
In Section~\ref{sec:isr_overview} we give an overview of the proof for Theorem~\ref{thm:main-2+3}. The proof of the 2-way part of Theorem~\ref{thm:main-2+3} appears in Section~\ref{sec:twoway_isr_pf}, and that of the 1-way part appears in Section~\ref{sec:oneway_isr_pf}. In Section~\ref{sec:1way_GHD_low_bds}, we prove a technical lemma needed in the proof of the 1-way part of Theorem~\ref{thm:main-2+3}.

\else
In this section, we sketch the proof of Theorem~\ref{thm:main-2+3}. To do so, we first provide, in Section~\ref{subsec:atomic_isr_ub}, some upper bounds on the ISR-CC of two ``atomic'' problems: Small Set Disjointness and $k$-Hamming Distance. 
The proof 
of the $2$-way part of Theorem~\ref{thm:main-2+3} is sketched in
Section~\ref{subsec:twoway_isr_pf}.
The proof of the $1$-way part shares many similarities with that of the $2$-way part and we defer it to the full version.
\fi


\subsection{ISR Protocols for Basic Problems}\label{sec:atomic_isr_ub}

In this section, we prove that ISR-CC and PSR-CC are polynomially related for some specific functions (note that this is stronger than Theorem~\ref{thm:main-2+3}, in the sense that the additive $O(\log \log n)$ factor is not required). In particular, we give ISR protocols for two basic problems: {\sc Small-Set-Intersection} (in Section~\ref{subsec:ssi_isr}) and {\sc Small-Hamming-Distance} (in Section~\ref{subsec:k_HD_isr}), such that the communication costs of these protocols are polynomially related to the respective PSR-CC of these functions. Our motivation in doing so is two-fold: firstly, to give techniques for designing efficient ISR protocols, and secondly because these protocols are at the heart of our proof of Theorem~\ref{thm:main-2+3}. In addition to these, we also give ISR-protocols for the class of {\em strongly permutation-invariant functions} which we describe in Section~\ref{subsec:strongly_perm_inv_subsec}.

\subsubsection{Small Set Intersection}\label{subsec:ssi_isr}

The {\sc Small-Set-Intersection} problem is defined as follows.
\begin{defn}[{\sc Small-Set-Intersection}]
$\SSI^n_{a,b} : \bit^n \times \bit^n \to \bbZ \union \set{?}$ is defined as follows,
$$\SSI^n_{a,b}(\bx, \by) = \infork{|\bx \land \by| & \text{if } |\bx| = a, |\by| = b \\ ? & \text{otherwise}}$$
Essentially, Alice is given $\bx \in \bit^n$ such that $|\bx| = a$, Bob is given $\by \in \bit^n$ such that $|\by| = b$, and they wish to compute $|\bx \land \by|$.
\end{defn}

\noindent The next two lemmas show that ISR-CC and PSR-CC are polynomially related for $\HD^n_k$ (for 1-way and 2-way models respectively).

\begin{lem}[1-way ISR Protocol for $\SSI^n_{a,b}$]\label{le:ssi_isr_prot}
Let $n, a, b \in \bbN$, such that, $a, b \le n/2$. Then,
$$\Omega(\max\set{a, \log b}) \le \R^{\oneway}(\SSI^n_{a,b}) \le \isr^{\oneway}_{\rho}(\SSI^n_{a,b}) = O(a \log(ab))$$
\end{lem}

\begin{proof}
We first describe the 1-way ISR-protocol for $\SSI^n_{a,b}$.

Let $\bx$ be Alice's string and $\by$ be Bob's string, with $a = |\bx|$ and $b = |\by|$. First, Alice and Bob use their correlated randomness to sample hash functions $h_A, h_B : [n] \rightarrow \set{0, 1}^r$ such that for any $i$, $h_A(i)$ and $h_B(i)$ are $\rho$-correlated strings, but for $i \ne j$, $h_A(i)$ and $h_B(j)$ are independent. Now, Alice sends $\setdef{h_A(i)}{x_i = 1}$, which Bob sees as $h_1, h_2, \dots, h_a$. Then, Bob computes the size of the set $\setdef{j \in \supp(\by)}{\Delta(h_B(j), h_i) \le \frac{1}{2} - \frac{\rho}{4} \text{ for some }i \in [a]}$ and outputs it.

By the Chernoff bound, we have that for any $i$, $\Delta(h_A(i), h_B(i)) \le \frac{1}{2}-\frac{\rho}{4}$ with probability $1 - 2^{-\Omega(r)}$. Also, for any $i \ne j$, $\Delta(h_A(i), h_B(j)) \ge \frac{1}{2}-\frac{\rho}{4}$ with probability $1 - 2^{-\Omega(r)}$. Thus, the probability that for every $i$ such that $x_i = y_i = 1$, $\Delta(h_A(i), h_B(i)) \le \frac{1}{2}-\frac{\rho}{4}$ and for every $i \ne j$ such that $x_i = y_j = 1$, $\Delta(h_A(i), h_B(i)) \ge \frac{1}{2}-\frac{\rho}{4}$ is at least $1 - ab 2^{-\Omega(r)}$ (by a union bound). Thus, with probability at least $1 - ab 2^{-\Omega(r)}$, Bob is able to correctly determine the exact value of $|\bx \wedge \by|$. Choosing $r = \Theta(\log(ab))$ yields a $1$-way ISR protocol with $O(a \log(ab))$ bits of communication from Alice to Bob.

The lower bound $\R^{\oneway}(\SSI^n_{a,b}) \ge \Omega(\max\set{a, \log b})$ will follow from Lemma~\ref{lem:1way_GHD_LB} which we proved in Section~\ref{sec:1way_GHD_low_bds}, because any protocol for $\SSI^n_{a,b}$ can be used to compute $\eGHD^n_{a,b,c,1}$, where we can choose $c$ to be anything, in particular, we choose $c \approx \max\set{a,b}$.
\end{proof}

\begin{lem}[2-way ISR Protocol for $\SSI^n_{a,b}$]\label{le:ssi_2way_isr_prot}
Let $n, a, b \in \bbN$. Let $a, b \le n/2$. Additionally, assume wlog that $a \le b$ (since the roles of Alice and Bob can be flipped). Then,
$$\Omega(\max\set{a, \log b}) \le \R(\SSI^n_{a,b}) \le \isr_{\rho}(\SSI^n_{a,b}) = O(a\log(ab))$$
\end{lem}
\begin{proof}
The ISR protocol is same as in proof of Lemma~\ref{le:ssi_2way_isr_prot}, with the difference that we flip the roles of Alice and Bob if $a \ge b$. The lower bound of $\R^{\oneway}(\SSI^n_{a,b}) \ge \Omega(\max\set{a, \log b})$ follows from Lemma~\ref{lem:main_low_bds_ghd} as any protocol for $\SSI^n_{a,b}$ also solves $\eGHD^n_{a,b,b,1}$.
\end{proof}

\begin{remark}\label{remark:gen_eq_smp}
The protocol given in proof of Lemma~\ref{le:ssi_isr_prot} can be viewed as a generalization of the protocol of \cite{bavarian2014role} for the {\sc Equality} function. More precisely, the {\sc Equality} function on $n$ bit strings is equivalent to the $\SSI^{2^n}_{1,1}$ problem. This is because Alice and Bob can keep a list of all $2^n$ elements of $\bit^n$ (e.g., in lexicographic order) and then view their input strings as subsets of cardinality $1$ of this list.
\end{remark}

\noindent We will repeatedly use the following corollary which follows from Lemma~\ref{le:ssi_isr_prot} by setting $a = 1$ and $b = 2^k$. Note that $\SSI^n_{1,2^k}$ is like the reverse direction of {\sc Sparse-Indexing}, in which Alice had a large set and Bob had a singleton set.

\begin{cor}[Reverse {\sc Sparse-Indexing}]\label{cor:isr_sparse_ind_cor}
$\forall \ n, k \in \bbN$, $\isr^{\oneway}(\SSI^n_{1,2^k}) = O(k)$.
\end{cor}

\subsubsection{Small Hamming Distance}\label{subsec:k_HD_isr}

The {\sc Small-Hamming-Distance} problem is defined as follows.

\begin{defn}[{\sc Small-Hamming-Distance}]
Let $n, k \in \bbN$ and $0 \le k \le n-1$. Then $\HD^n_k : \bit^n \times \bit^n \to \bit$ is defined as follows,
$$\HD^n_k(\bx, \by) = \infork{1 & \text{if } \Delta(\bx,\by) \le k \\ 0 & \text{if } \Delta(\bx,\by) > k}$$
Essentially, Alice is given $\bx \in \bit^n$ and Bob is given $\by \in \bit^n$ and they wish to distinguish between the cases $\Delta(\bx,\by) \le k$ and $\Delta(\bx,\by) > k$.
\end{defn}

\noindent The following lemma shows that ISR-CC and PSR-CC are polynomially related for $\HD^n_k$ (for both the 1-way and 2-way models).

\begin{lem}[ISR Protocol for {\sc Small-Hamming-Distance}]\label{le:main_isr_up_bd_k_HD}
Let $n, k \in \bbN$. Additionally, assume wlog that $k \le n/2$ (since Bob can flip his input, and thus computing $\HD^n_k$ is equivalent to computing $\HD^n_{n-k}$). Then,
$$\Omega(k) \le \R(\HD^n_k) \le \isr^{\oneway}(\HD^n_k) \le O(k^2)$$
\end{lem}

\ifnum\focs=1
Our protocol builds on a recent $1$-way ISR protocol for estimating inner products of real unit vectors from \cite{CGMS_ISR}. The proof of Lemma~\ref{le:main_isr_up_bd_k_HD} appears in the full version.
\else

\noindent In order to prove Lemma~\ref{le:main_isr_up_bd_k_HD}, we will use the following protocol (from \cite{CGMS_ISR}) twice, namely in the proofs of Lemmas~\ref{le:isr_small_vs_huge_HD} and \ref{le:isr_small_vs_moderate}.

\begin{lem}[ISR protocol for {\sc Gap-Inner-Product} \cite{CGMS_ISR}]\label{le:isr_CGMS_up_bd}
Let $-1 \le s < c \le 1$ be real numbers. Assume that Alice is given a vector $\bu \in \mathbb{R}^n$ such that $|| \bu ||_2 = 1$, and that Bob is given a vector $\bv \in \mathbb{R}^n$ such that $|| \bv ||_2 = 1$. Then, there is a $1$-way ISR protocol that distinguishes between the cases $\inangle{\bu,\bv} \geq c$ and $\inangle{\bu,\bv} \le s$ using $O(1/(c-s)^2)$ bits of communication from Alice to Bob.
\end{lem}

\begin{lem}\label{le:isr_small_vs_huge_HD}
Assume that $k<n/20$. Then, there is a $1$-way ISR protocol that distinguishes between the cases $\Delta(x,y) \le k$ and $\Delta(x,y) > n/10$ using $O(1)$ bits of communication.
\end{lem}

\begin{proof}
Let Alice construct the vector $\bu \in \mathbb{R}^n$ by setting $\bu_i = (-1)^{\bx_i}/\sqrt{n}$ for every $i \in [n]$ and let Bob construct the vector $\bv \in \mathbb{R}^n$ by setting $\bv_i = (-1)^{\by_i}/\sqrt{n}$ for every $i \in [n]$. Then, we have that $|| \bu ||_2 = || \bv ||_2 = 1$. Furthermore, $\langle \bu , \bv \rangle = 1-2\Delta(\bx,\by)/n$. Therefore, $\Delta(\bx,\by) \le k$ implies that $\langle \bu , \bv \rangle \geq 1-2k/n$ and $\Delta(\bx,\by) > n/10$ implies that $\langle \bu , \bv \rangle < 4/5$. Setting $c: = 1-2k/n$ and $s:=  4/5$ and using the assumption that $k < n/20$, Lemma~\ref{le:isr_CGMS_up_bd} yields a $1$-way ISR protocol with $O(1)$ bits of communication from Alice to Bob (since $(c-s) \ge 1/10$).
\end{proof}

\begin{lem}\label{le:isr_small_vs_moderate}
Assume that $k < n/20$. Then, there is a $1$-way ISR protocol that distinguishes between the cases $\Delta(x,y) \le k$ and $k<\Delta(x,y)\le n/10$ using $O(k^2)$ bits of communication.
\end{lem}

\begin{proof}
As in the proof of Lemma~\ref{le:isr_small_vs_huge_HD}, we let Alice and Bob construct unit vectors $\bu, \bv \in \mathbb{R}^n$ by setting $\bu_i = (-1)^{\bx_i}/\sqrt{n}$ and $\bv_i = (-1)^{\by_i}/\sqrt{n}$ (respectively) for every $i \in [n]$. Then, $\langle \bu , \bv \rangle = 1-2\Delta(\bx,\by)/n$. Let $t:= n/(10k)$. Alice tensorizes her vector $t$ times to obtain the vector $\bu^{\otimes t} \in \mathbb{R}^{n^t}$, namely, for every $i_1,i_2,\dots,i_t \in [n]$, she sets $\bu^{\otimes t}_{(i_1,i_2,\dots,i_t)} = \prod\limits_{j=1}^{t} u_{i_j}$. Similarly, Bob tensorizes his vector $t$ times to obtain the vector $\bv^{\otimes t} \in \mathbb{R}^{n^t}$. Observe that $|| \bu^{\otimes t} ||_2 = || \bu ||_2^t = 1$, $|| \bv^{\otimes t} ||_2 = || \bv ||_2^t = 1$ and
$$ \langle \bu^{\otimes t} , \bv^{\otimes t} \rangle = \langle \bu , \bv \rangle^t =  (1-2\Delta(\bx,\by)/n)^t.$$
Therefore, $\Delta(\bx,\by) \le k$ implies that $\langle \bu^{\otimes t} , \bv^{\otimes t} \rangle \geq (1-2k/n)^t := c$, and $k<\Delta(x,y)\le n/10$ implies that $\langle \bu^{\otimes t} , \bv^{\otimes t} \rangle \le (1-2(k+1)/n)^t := s$. The inner product gap $c-s$ is at least
\begin{eqnarray*}
\inparen{1-\frac{2k}{n}}^t - \inparen{1-\frac{2(k+1)}{n}}^t &=& \inparen{1-\frac{2k}{n}}^t \insquare{1- \inparen{1-\frac{1}{n/2-k}}^t}\\
&\geq& \inparen{1- \frac{2kt}{n}}\inparen{1-e^{-t/(n/2-k)}}\\ 
&\ge& \frac{4}{5}\inparen{1-e^{-2/(9k)}} \quad \insquare{\because t = n/10k \sAND k < n/20}\\ 
&\ge& \Omega\inparen{\frac{1}{k}}
\end{eqnarray*}
where the first inequality above follows from the fact that $(1+x)^r \le e^{rx}$ for every $x,r \in \mathbb{R}$ with $r>0$, as well as the fact that $(1-x)^r \geq 1-xr$ for every $0 \le x \le 1$ and $r \geq 1$. Moreover, the last equality above follows from the fact that $(1-e^{-x})/x \to 1$ as $x \to 0$. Therefore, applying Lemma~\ref{le:isr_CGMS_up_bd} with $c-s = \Omega(1/k)$ yields a $1$-way ISR protocol with $O(k^2)$ bits of communication from Alice to Bob.
\end{proof}

\noindent We are now ready to prove of Lemma~\ref{le:main_isr_up_bd_k_HD}.

\begin{proof}[Proof of Lemma~\ref{le:main_isr_up_bd_k_HD}]
Assume without loss of generality that $k < n/20$, since otherwise, Alice can simply send her entire input ($n$ bits) to Bob, requiring only $O(k)$ communication. Run the protocols from Lemmas~\ref{le:isr_small_vs_huge_HD} and \ref{le:isr_small_vs_moderate} in sequence\footnote{More precisely, we first repeat each of the two protocols a constant number of times and take a majority-vote of the outcomes. This allows us to reduce the error probability to a small enough constant and thereby apply a union bound.} and declare that $\Delta(x,y) \le k$ if and only if both protocols say so; otherwise, declare that $\Delta(x,y) > k +1$. This gives a $1$-way ISR protocol with $O(k^2)$ bits of communication from Alice to Bob.

The lower bound $\R(\HD^n_k) \ge \Omega(k)$ follows from Lemma~\ref{lem:main_low_bds_ghd} as any protocol the computes $\HD^n_k$ can be used to compute $\eGHD^n_{k,k,k,1}$ as well.
\end{proof}

\subsubsection{Strongly Permutation-Invariant functions}\label{subsec:strongly_perm_inv_subsec}
In this section, we show that the ISR-CC is polynomially related to the PSR-CC -- without any additive dependence on $n$ -- for a natural subclass of permutation-invariant functions that we call ``strongly permutation-invariant functions''. We point out that this section is not needed for proving Theorem~\ref{thm:main-2+3}, but we include it because it highlights some of the proof ideas that we eventually use. We start by defining strongly permutation-invariant functions.

\begin{defn}[(Total) Strongly Permutation-Invariant functions]
A (total) function $f : \bit^n \times \bit^n \to \bit$ is {\em strongly permutation-invariant} if there exists a symmetric function $\sigma : \bit^n \to \bit$ and a function $h:\bit^2 \to \bit$ such that for every $\bx, \by \in \bit^n$, $$f(\bx, \by) = \sigma(h(x_1, y_1), h(x_2, y_2), \cdots, h(x_n, y_n))$$
\end{defn}

\noindent Note that strongly permutation-invariant functions include as subclasses, (AND)-symmetric functions (studied, e.g.,  by \cite{buhrman2001communication, razborov2003quantum,sherstov2011unbounded}) and XOR-symmetric functions (studied, e.g., by \cite{zhang2009communication}).

\noindent The following theorem shows that ISR-CC of any strongly permutation-invariant function is polynomially related to its PSR-CC, with no dependence on $n$.

\begin{theorem}\label{thm:strong_pi}
For any total strongly permutation-invariant function $f : \bit^n \times \bit^n \to \bit$, if $\R(f) = k$ then,
\begin{eqnarray*}
\isr(f) &\le& \Otilde(k^2)\\
\isr^{\oneway}(f) &\le& \Otilde(k^3)
\end{eqnarray*}
\end{theorem}

\begin{proof}
Depending on $h$, any such function depends only on the sum of some subset of the quantities $\set{|\bx \wedge \by|, |\bx \wedge \lnot \by|, |\lnot \bx \wedge \by|, |\lnot \bx \wedge \lnot \by|}$. There are three main cases to consider (the remaining cases being similar to these three):

\begin{enumerate}
\item[(i)] {\em $f$ depends only on $|\bx \wedge \by| + |\bx \wedge \lnot \by|$:} In this case, $f$ depends only on $|\bx|$, and hence $\R(f)$, $\isr(f)$ and $\isr^{\oneway}(f)$ are all $1$.

\item[(ii)] {\em $f$ depends only on $|\bx \wedge \by| + |\lnot \bx \wedge \lnot \by|$:} In this case, $f$ depends only on $|\bx \oplus \by|$. Let $\R(f) = k$, and suppose $i$ is such that $f(\bx, \by) = 0$ for $|\bx \oplus \by| = i-1$, and $f(\bx, \by) = 1$ for $|\bx \oplus \by| = i+1$. If $i \le n/2$ then any protocol to compute $f$ can be used to compute $\eGHD^n_{i,i,i,1}$. Applying Lemma~\ref{lem:main_low_bds_ghd} we get that, $k = \R(f) \ge \R(\eGHD^n_{i,i,i,1}) \ge \IC(\eGHD^n_{i,i,i,1}) \ge i/C$. If $i > n/2$, then any protocol to compute $f$ can be used to compute $\eGHD^n_{i,i,i,1}$. Applying Lemma~\ref{lem:main_low_bds_ghd} again, we get that, $k = \R(f) \ge \R(\eGHD^n_{i, n-i, i, 1}) \ge \IC(\eGHD^n_{i,n-i,i,1}) \ge \Omega(n-i)$. Thus, we get that for any such $i$, it must be the case that either $i \le Ck$ or $i \ge n-Ck$.

Alice and Bob now use the $1$-way ISR protocol given by Lemma~\ref{le:main_isr_up_bd_k_HD} to solve $\HD^n_i$ for every $i$ such that $i \le Ck$ or $i \ge n-Ck$, and for each such problem, they repeat the protocol $O(\log{k})$ times to make the error probability down to $O(1/k)$. This yields a $1$-way ISR protocol with $\Otilde(k^3)$ bits of communication from Alice to Bob. This protocol can be modified into a $2$-way ISR protocol with only $\Otilde(k^2)$ bits of communication by letting Alice and Bob binary-search over the $O(k)$ Hamming distance problems that they need to solve, instead of solving all of them in parallel.

\item[(iii)] {\em $f$ depends only on $|\bx \wedge \by|$:} Suppose $j$ is such that $f(\bx, \by) = 0$ when $|\bx \wedge \by| = j$ and $f(\bx, \by) = 1$ when $|\bx \wedge \by| = j+1$. Then, if we restrict to only $\bx$ and $\by$ such that $|\bx| = |\by| = (n+2j)/3$, then any protocol to compute $f$ can be used to compute $\eGHD^n_{a, a, c, 1}$, where $a = (n+2j)/3$ and $c=(2n-2j)/3$. Applying Lemma~\ref{lem:main_low_bds_ghd} we get that, $k = \R(f) \ge \R(\eGHD^n_{a,a,c,1}) \ge \IC(\eGHD^n_{a,a,c,1}) \ge 2(n-j)/3C$. This implies that $j \geq n- 2Ck$. In particular, we deduce that there are at most $O(k)$ such values of $j$. On input pair $(\bx,\by)$, Alice checks whether $|\bx| \geq n-2Ck$ and Bob checks whether $|\by| \geq n-2Ck$. If one of these inequalities fail, then it is the case that $|\bx \land \by| < n-2Ck$ and the function value can be deduced. Suppose that $|\bx| \geq n-100k$ and $|\by| \geq n-100k$. In this case, $|\bx|$ takes one of $O(k)$ possibilities and hence Alice can send $|\bx|$ to Bob using $O(\log{k})$ bits. At this point, Bob knows both $|\bx|$ and $|\by|$. Thus, using the identity $\Delta(\bx,\by) = |\bx| + |\by| - 2|\bx \land \by|$, the problem gets reduced to a collection $O(k)$-Hamming Distance problems as in case (ii) above. The same protocols as in case (ii) imply a $1$-way ISR protocol with $\Otilde(k^3)$ bits of communication from Alice to Bob, and a $2$-way ISR protocol with $\Otilde(k^2)$ bits of communication.\qedhere
\end{enumerate}
\end{proof}

\fi


\subsection{Overview of Proofs} \label{sec:isr_overview}
The proofs of the $1$-way and $2$-way parts of Theorem~\ref{thm:main-2+3} follow the same general framework, which we describe next. Let $(\bx,\by)$ be the input pair, $a:=|\bx|$ and $b:= |\by|$. We partition the $(a,b)$-plane into a constant number of regions such that:
\begin{enumerate}
\item[(i)] Using a small amount of communication, Alice and Bob can distinguish in which region their combined input lies.
\item[(ii)] For each of these regions, there is an ISR protocol with small communication that computes the function value on any input in this region.
\end{enumerate}
Some of the region-specific protocols (in (ii) above) will be based on low-communication ISR protocols for two ``atomic'' problems: {\sc Small-Set-Intersection} and {\sc Small-Hamming-Distance} described in Section~\ref{sec:atomic_isr_ub}.

We point out again that both of our protocols for {\sc Small-Set-Intersection} and {\sc Small-Hamming-Distance} have ISR-CC that is polynomial in the underlying PSR-CC, which is crucial for our purposes. In particular, one cannot instead use the generic exponential simulation of \cite{CGMS_ISR}.

The additive $O(\log\log{n})$ factor in the ISR-CC upper bound of Theorem~\ref{thm:main-2+3} is due to the fact that for one region (other than the ones mentioned above), we show that in order for Alice and Bob to compute the function value $f(\bx,\by)$, it is enough that they compute some other low PSR-CC function $f'(|\bx|, |\by|)$ of the {\em Hamming weights} of $(\bx, \by)$. Since the Hamming weight of an $n$-bit string can be expressed using $O(\log{n})$ bits, we have effectively reduced the ``dimension'' of the problem from $n$ to $O(\log{n})$. At this point, we can apply Newman's theorem (Theorem~\ref{thm:newman}) to obtain a private-coin protocol computing $f'(|\bx|,|\by|)$ (and hence $f(\bx,\by)$) while increasing the communication cost by at most an additive $O(\log\log{n})$.

\subsection{2-way ISR Protocol for Permutation-Invariant Functions}\label{sec:twoway_isr_pf}
In this section, we prove the $2$-way part of Theorem~\ref{thm:main-2+3}. We again use the measure $m(f)$ (introduced in Definition~\ref{def:measure_2way_partial}) when restricted to \emph{total} functions. For the sake of clarity, we describe the resulting specialized expression of $m(f)$ again in the following proposition.

\begin{proposition}[Measure $m(f)$ for total functions] \label{prop:measure_2way_total}
Given a total permutation-invariant function $f : \bit^n \times \bit^n \to \bit$, and integers $a, b$, s.t. $0 \le a, b \le n$, let $h_{a,b} : \set{0, 1, \cdots, n} \to \set{0,1,?}$ be the function given by $h_{a,b}(d) = f(\bx, \by)$ if there exist $\bx$, $\by$ with $|\bx| = a$, $|\by| = b$, $\Delta(\bx, \by) = d$ and $?$ otherwise. (Note. by permutation invariance of $f$, $h_{a,b}$ is well-defined.) Let $\calJ(h_{a,b})$ be the set of {\em jumps} in $h_{a,b}$, defined as follows,
$$\calJ(h_{a,b}) \quad \defeq \quad \setdef{c}{\inmat{h_{a,b}(c-1) \ne h_{a,b}(c+1) \\ h_{a,b}(c-g), \ h_{a,b}(c+g) \in \bit}}$$
Then, we define $m(f)$ as follows.
$$m(f) \quad \defeq \quad \frac{1}{C} \cdot \max_{\substack{a, b \in [n] \\ (c,g) \in \calJ(h_{a,b})}} \max\set{\min\set{a,b,c,n-a,n-b,n-c}, \log\inparen{\min \set{c, n-c}}}$$
where $C$ is a suitably large constant (which does not depend on $n$).
\end{proposition}

\noindent We will now prove the following theorem, which immediately implies the 2-way part of Theorem~\ref{thm:main-2+3}.
\begin{theorem}\label{thm:isr_thm_2_way}
Let $f : \bit^n \times \bit^n \rightarrow \bit$ be any (total) permutation-invariant function. Then,
$$m(f) \le \R(f) \le \isr(f) \le \Otilde(m(f)^3) + \R(f) + O(\log \log n)$$
\end{theorem}

\ifnum\focs=1
\noindent We now sketch the proof of Theorem~\ref{thm:isr_thm_2_way}. The full proof appears in the full version.
\begin{proof-sketch}
\else
\begin{proof}
\fi
Any protocol to compute $f$ also computes $\eGHD^n_{a,b,c,g}$ for any $a$, $b$ and any jump $c \in \calJ(h_{a,b})$. Consider the choice of $a$, $b$ and a jump $c \in \calJ(h_{a,b})$ such that the lower bound obtained on $\IC(\eGHD^n_{a,b,c,1})$ through Lemma~\ref{lem:main_low_bds_ghd} is maximized, which by definition is equal to $m(f)$ (after choosing an appropriate $C$ in the definition of $m(f)$). Thus, we have $m(f) \le \IC(\eGHD^n_{a,b,c,g}) \le \IC(f) \le \R(f)$.

Let $k \defeq m(f)$. The main part of the proof is to show that $\isr(f) \le \Otilde(k^3) + \R(f) + O(\log \log n)$. We first divide the input space into a constant number of regions, such that Alice can send $O(\log{k})$ number of bits to Bob with which he can decide in which of the regions does their combined input lie (with high probability). Thus, once we break down the input space into these regions, it will suffice to give $2$-way protocols with small ISR-CC for computing the function over each of these regions; as Alice and Bob can first determine in which region their combined input lies, and then run the corresponding protocol to compute the function value.

\ifnum\focs=0
Let $a = |\bx|$ and $b = |\by|$. We divide the $(a,b)$-plane into $4$ regions, (I), (II), (III) and (IV),  based on the values of $a$ and $b$ as follows. First let $A = \min\set{a, n-a}$ and $B = \min\set{b, n-b}$. Then the regions are given by,

\begin{enumerate}
\item[(I)] ($A \le Ck$ and $B \le 2^{Ck}$) or ($A \le 2^{Ck}$ and $B \le Ck$)
\item[(II)] ($A \le Ck$ and $B > 2^{Ck}$) or ($A > 2^{Ck}$ and $B \le Ck$)
\item[(III)] $A > Ck$ and $B > Ck$ and $|A-B| < Ck$
\item[(IV)] $A > Ck$ and $B > Ck$ and $|A-B| \geq Ck$
\end{enumerate}

\noindent where $C$ comes from Definition~\ref{def:measure_2way_partial}. Note that regions (I), (II), (III) and (IV) form a {\em partition} of the $(a,b)$-plane. This division is shown pictorially in Figure~\ref{fig:division-2}.

First, note that if $|\bx| > n/2$, then Alice can flip all her input bits and convey that she did so to Bob using one bit of communication. Similarly, if $|\by| > n/2$, then Bob can flip all his input bits and convey that he did so to Alice using one bit of communication. After these flipping operations, Alice and Bob will look at the appropriately modified version of $f$ based on who all flipped their input. Note that flipping all the bits of Alice and/or Bob preserves the permutation-invariance of the function. We will henceforth assume w.l.o.g. that $a = |\bx| \le n/2$ and $b = |\by| \le n/2$. [This is also the reason that the regions described above are in terms of $A = \min\set{a, n-a}$ and $B = \min\set{b, n-b}$ and $A, B \le n/2$.]

Next, we show that determining the region to which the input pair $(\bx, \by)$ belongs can be done using $O(\log{k})$ bits of (ISR) communication from Alice to Bob. First, Alice will send Bob two bits indicating whether $a \le Ck$ and whether $a \le 2^{Ck}$ respectively. With this information Bob can determine in which of regions \{(I), (II), (III) $\cup$ (IV)\} the combined input lies. To distinguish between regions (III) and (IV), Alice and Bob can first check whether $|a-b| < 100k$ by setting up an instance of {\sc Sparse-Indexing}. Namely, Alice will translate the value $a$ into a string $s_a$ where $s_a(i) = 1$ iff $i=a$. And Bob will translate $b$ into a string $s_b$ such that $s_b(i) = 1$ iff $b-Ck < i < b+Ck$. This is an instance of {\sc Sparse-Indexing} which can be solved with $O(\log k)$ bits of 
\ifnum\focs=0
ISR-CC, by Corollary~\ref{cor:isr_sparse_ind_cor}. 
\else
ISR-CC.
\fi

\ifnum\focs=1
In case (I), the problem gets reduced to a Small Set Intersection problem and we use 
Lemma~\ref{le:ssi_isr_prot} to give a protocol handling this case. 
In case (II), there turns out to be no dependence on $\Delta(\bx,\by)$ and Alice can send $|\bx|$ to Bob with $O(\log k)$
bits and Bob can compute the function.
In case (III), the problem turns out to be an instance of Small Hamming Distance, and we can use 
Lemma~\ref{le:main_isr_up_bd_k_HD} to determine this distance.
Finally, in Case (IV), again there is no dependence on the Hamming Distance, but here Alice cannot communicate $|\bx|$.
But we can use Newman's theorem to compute the function $h_{a,b}(\cdot)$ using $k + O(\log\log n)$ bits.
(There is a randomized $k$ bit protocol for $h$ since there is one for $f$, and the protocol's randomness can be reduced
since the inputs are only in the range $[n]$.)
\else
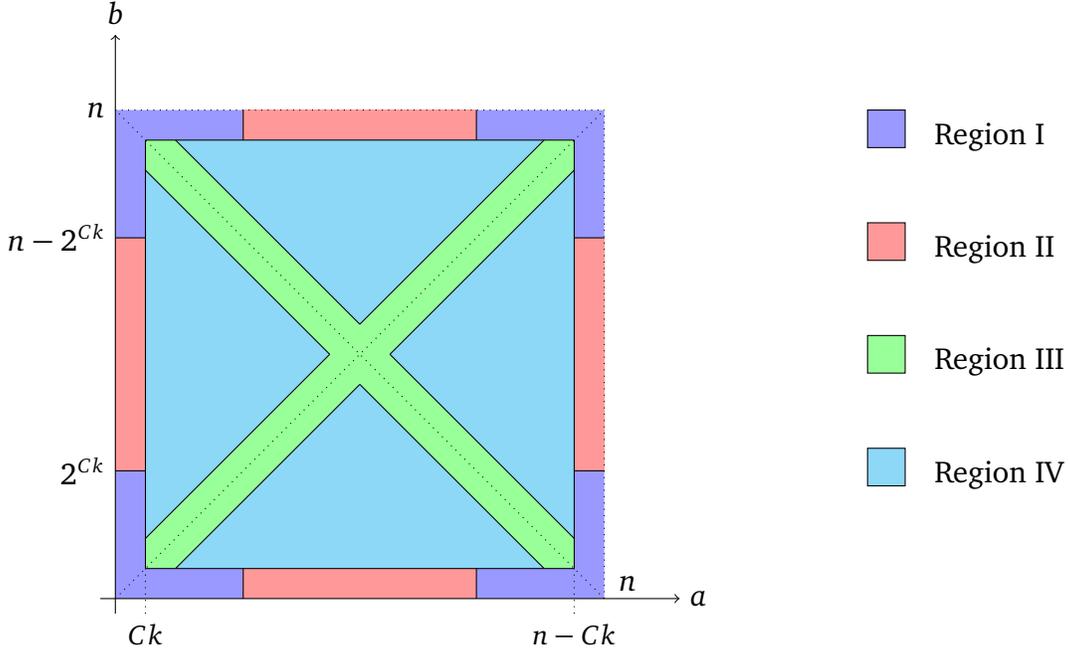
\begin{figure}
\begin{center}
\begin{tikzpicture}
\def \n {6.5};
\def \ex {1};
\def \smex {0.1};
\def \k {0.4};
\def \expk {1.7};

\def \index {\n+4};
\def \len {0.5};
\def \hgt {1.5};

\def \bgt{40}
\def \colA {blue!\bgt};
\def \colB {red!\bgt};
\def \colC {green!\bgt};
\def \colD {cyan!\bgt};

\coordinate (A0) at (0,0);
\coordinate (A1) at (\n,0);
\coordinate (A2) at (\n,\n);
\coordinate (A3) at (0,\n);

\coordinate (K01) at (2*\k,\k);
\coordinate (K02) at (\k,\k);
\coordinate (K03) at (\k,2*\k);
\coordinate (K11) at (\n-2*\k,\k);
\coordinate (K12) at (\n-\k,\k);
\coordinate (K13) at (\n-\k,2*\k);
\coordinate (K21) at (\n-2*\k,\n-\k);
\coordinate (K22) at (\n-\k,\n-\k);
\coordinate (K23) at (\n-\k,\n-2*\k);
\coordinate (K31) at (2*\k,\n-\k);
\coordinate (K32) at (\k,\n-\k);
\coordinate (K33) at (\k,\n-2*\k);

\coordinate (E01) at (\expk,0);
\coordinate (E02) at (\expk,\k);
\coordinate (E03) at (\k,\k);
\coordinate (E04) at (\k,\expk);
\coordinate (E05) at (0,\expk);
\coordinate (E11) at (\n-\expk,0);
\coordinate (E12) at (\n-\expk,\k);
\coordinate (E13) at (\n-\k,\k);
\coordinate (E14) at (\n-\k,\expk);
\coordinate (E15) at (\n-0,\expk);
\coordinate (E21) at (\n-\expk,\n);
\coordinate (E22) at (\n-\expk,\n-\k);
\coordinate (E23) at (\n-\k,\n-\k);
\coordinate (E24) at (\n-\k,\n-\expk);
\coordinate (E25) at (\n,\n-\expk);
\coordinate (E31) at (\expk,\n);
\coordinate (E32) at (\expk,\n-\k);
\coordinate (E33) at (\k,\n-\k);
\coordinate (E34) at (\k,\n-\expk);
\coordinate (E35) at (0,\n-\expk);

\coordinate (M01) at (\n/2,\n/2-\k);
\coordinate (M12) at (\n/2+\k,\n/2);
\coordinate (M23) at (\n/2,\n/2+\k);
\coordinate (M30) at (\n/2-\k,\n/2);

\draw[draw=none, fill=\colA] (A0) -- (E01) -- (E02) -- (E03) -- (E04) -- (E05) -- cycle;
\draw[draw=none, fill=\colA] (A1) -- (E11) -- (E12) -- (E13) -- (E14) -- (E15) -- cycle;
\draw[draw=none, fill=\colA] (A2) -- (E21) -- (E22) -- (E23) -- (E24) -- (E25) -- cycle;
\draw[draw=none, fill=\colA] (A3) -- (E31) -- (E32) -- (E33) -- (E34) -- (E35) -- cycle;

\draw[draw=none, fill=\colB] (E05) -- (E04) -- (E34) -- (E35) -- cycle;
\draw[draw=none, fill=\colB] (E01) -- (E02) -- (E12) -- (E11) -- cycle;
\draw[draw=none, fill=\colB] (E15) -- (E14) -- (E24) -- (E25) -- cycle;
\draw[draw=none, fill=\colB] (E21) -- (E22) -- (E32) -- (E31) -- cycle;

\draw[draw=none, fill=\colC] (K03) -- (K02) -- (K01) -- (M01) -- (K11) -- (K12) -- (K13) -- (M12) -- (K23) -- (K22) -- (K21) -- (M23) -- (K31) -- (K32) -- (K33) -- (M30) -- cycle;

\draw[draw=none, fill=\colD] (K01) -- (M01) -- (K11) -- cycle;
\draw[draw=none, fill=\colD] (K13) -- (M12) -- (K23) -- cycle;
\draw[draw=none, fill=\colD] (K21) -- (M23) -- (K31) -- cycle;
\draw[draw=none, fill=\colD] (K33) -- (M30) -- (K03) -- cycle;

\node[right] (a) at (\n+\ex,0) {$a$};
\node[above] (b) at (0,\n+\ex) {$b$};

\draw[->] (-\ex/5,0) -- (a);
\draw[->] (0,-\ex/5) -- (b);

\draw (E03) -- (E13);
\draw (E13) -- (E23);
\draw (E23) -- (E33);
\draw (E33) -- (E03);

\draw (E01) -- (E02);
\draw (E11) -- (E12);
\draw (E21) -- (E22);
\draw (E31) -- (E32);
\draw (E04) -- (E05);
\draw (E14) -- (E15);
\draw (E24) -- (E25);
\draw (E34) -- (E35);

\draw (K01) -- (M01);
\draw (K11) -- (M01);
\draw (K13) -- (M12);
\draw (K23) -- (M12);
\draw (K21) -- (M23);
\draw (K31) -- (M23);
\draw (K33) -- (M30);
\draw (K03) -- (M30);

\draw[dotted] (A0) -- (A2);
\draw[dotted] (A1) -- (A3);
\draw[dotted] (A3) -- (A2);
\draw[dotted] (A1) -- (A2);

\node[below] at (\k,-2*\smex) {\small $Ck$} edge[dotted] (E03);
\node[below] at (\n-\k,-2*\smex) {\small $n-Ck$} edge[dotted] (E13);
\node[right] at (\n+0.5*\smex,2*\smex) {$n$};
\node[left] at (0,\n) {$n$};
\node[left] at (0,\expk) {$2^{Ck}$};
\node[left] at (0,\n-\expk) {$n-2^{Ck}$};

\draw[fill=\colD] (\index, \hgt) rectangle (\index-\len, \hgt+\len);
\draw[fill=\colC] (\index, 2*\hgt) rectangle (\index-\len, 2*\hgt+\len);
\draw[fill=\colB] (\index, 3*\hgt) rectangle (\index-\len, 3*\hgt+\len);
\draw[fill=\colA] (\index, 4*\hgt) rectangle (\index-\len, 4*\hgt+\len);

\node[right] at (\index+\len/2, \hgt+\len*0.3) {Region IV};
\node[right] at (\index+\len/2, 2*\hgt+\len*0.3) {Region III};
\node[right] at (\index+\len/2, 3*\hgt+\len*0.3) {Region II};
\node[right] at (\index+\len/2, 4*\hgt+\len*0.3) {Region I};

\end{tikzpicture}
\caption{Division of the $(a,b)$ plane into four regions in the $2$-way setting}
\label{fig:division-2}
\end{center}
\end{figure}


We now show how to compute the value of the function $f$ in each of the $4$ individual regions (I), (II), (III) and (IV) using ISR-CC of at most $\Otilde(k^3) + \R(f) + O(\log \log n)$ bits. Since $f$ is a permutation-invariant function, we use the following $2$ interchangeable representations of $f$ using Observation~\ref{obs:interchang_rep},
$$f(\bx, \by) = g(|\bx|, |\by|, |\bx \land \by|) = h(|\bx|, |\by|, \Delta(\bx, \by))$$

\begin{enumerate}

\item[(I)] (Main idea: {\sc Small-Set-Intersection}) We have that either ($a \le Ck$ and $b \le 2^{Ck}$) or ($a \le 2^{Ck}$ and $b \le Ck$). Since we can interchange the roles of Alice and Bob if required, we can assume w.l.o.g. that $a \le Ck$ and $b \le 2^{Ck}$. In this case, Alice first sends the value of $a = |\bx|$ to Bob. They can then apply the protocol from Lemma~\ref{le:ssi_isr_prot} in order to compute $|\bx \land \by|$ using $O(a\log(ab)) = O(k^2)$ bits of $1$-way ISR communication from Alice to Bob. Hence, Bob can determine $|\bx \land \by|$ correctly with high probability, and hence deduce $g(|\bx|, |\by|, |\bx \land \by|) = f(\bx, \by)$.

\item[(II)] (Main idea: {\em No dependence on $\Delta(\bx, \by)$}) We have that either ($a \le Ck$ and $b > 2^{Ck}$) or ($a > 2^{Ck}$ and $b \le Ck$). Since we can interchange the roles of Alice and Bob if required, we can assume w.l.o.g. that $a \le Ck$ and $b > 2^{Ck}$. Then, the definition of the measure $m(f)$ implies that for this range of values of $a$ and $b$, the function $h$ cannot depend on $\Delta(\bx, \by)$ (because in this case $\Delta(\bx, \by) \ge (b-a)$). Since $h$ depends only on $|\bx|$ and $|\by|$, Alice can simply send the value of $a$ (which takes only $O(\log k)$ bits), with which Bob can compute $h(a, b, c)$ for any valid $c$, that is, $h_{a,b}(c) \ne \ ?$.

\item[(III)] (Main idea: {\sc Small-Hamming-Distance}) We have that $|a-b| < 100k$. In this case, Alice sends the value of $a~(\mod 2Ck)$ to Bob, requiring $O(k)$ 1-way communication. Since Bob knows $b$, and so he can figure out the exact value of $a$. Next, they need to determine the Hamming distance $\Delta(\bx, \by)$. The definition of our measure $m(f)$ (along with the fact that $k:=m(f)$) implies that if $h(a,b,c-1) \ne h(a,b,c+1)$, then $c$ must be either $\le Ck$ or $\geq n-Ck$. That is, given $a$ and $b$, $h(a, b, c)$ must equal a constant for all valid $c$ such that $Ck < c < n - Ck$.

Since Bob knows both $a$ and $b$ exactly, Alice and Bob can run the 1-way ISR-protocol for $\HD^n_i$ (from Lemma~\ref{le:main_isr_up_bd_k_HD}) for every $i < Ck$ and $i > n-Ck$. This requires $\Otilde(k^3)$ communication.\footnote{The ISR-protocol for $\HD^n_i$ takes $O(i^2) \le O(k^2)$ communication each. But we can only afford to make error with probability $O(1/k)$, and thus for sake of amplification, the overall communication is $O(k^3 \log k)$.} If the Hamming distance is either $\le Ck$ or $\geq n-Ck$, then Bob can determine it exactly and output $h(a,b,c)$. If the Hamming distance is not in this range, then the Hamming distance does not influence the value of the function, in which case Bob can output $h(a, b, c)$ for any valid $c$ such that $Ck < c < n - Ck$.

\item[(IV)] (Main idea: {\em Newman's theorem on $(a,b)$}) We claim that for this range of values of $a$ and $b$, the function $h(a, b, c)$ cannot depend on $c$. Suppose for the sake of contradiction that the function depends on $c$. Hence, there exists a value of $c$ for which $h(a, b, c-1) \ne h(a, b, c+1)$. Since $|a - b| \ge Ck$, we have that $c \ge Ck$. And since $|n-a-b| \geq Ck$, we also have that $c \le \min\set{a+b, 2n-a-b} \le n - Ck$. Thus, we get that $Ck \le c \le n-Ck$, which contradicts that $m(f) = k$ (see Proposition~\ref{prop:measure_2way_total}).

Since $f(\bx, \by)$ only depends on $|\bx|$ and $|\by|$ in this region, we have converted the problem which depended on inputs of size $n$, into a problem which depends on inputs of size $O(\log{n})$ only. Since the original problem had a PSR-protocol with $\R(f)$ bits of communication, applying Newman's theorem (Theorem~\ref{thm:newman}), we conclude that, with private randomness itself, the problem can be solved using $\R(f) + O(\log \log n)$ bits of $2$-way communication.\qedhere
\end{enumerate}
\fi
\ifnum\focs=1
\end{proof-sketch}
\else
\end{proof}
\fi

\begin{note}
We point that the proof of Theorem~\ref{thm:isr_thm_2_way} shows, more strongly, that for any function $G(.)$, the following two statements are equivalent:
\begin{enumerate}
\item[(1)] Any total function $f$ with $\R(f) = k$ has $\isr(f) \le \poly(k) + G(n)$
\item[(2)] Any total permutation-invariant function $f$ with $\R(f) = k$ has $\isr(f) \le \poly(k) + G(O(\log n))$.
\end{enumerate}
Newman's theorem (Theorem~\ref{thm:newman}) implies that (1) holds for $G(n) = O(\log{n})$, and hence we have that (2) also holds for $G(n) = O(\log{n})$; thereby yielding the bound implied by Theorem~\ref{thm:isr_thm_2_way}. On the other hand, improving the $O(\log \log n)$ term in Theorem~\ref{thm:isr_thm_2_way} to, say $o(\log \log n)$ will imply that for all total functions we can have $G(n)$ in (1) to be $o(\log n)$. We note that currently such a statement is unknown.
\end{note}


\subsection{1-way ISR Protocol for Permutation-Invariant Functions}\label{sec:oneway_isr_pf}
In this section, we prove the 1-way part of Theorem~\ref{thm:main-2+3}. On a high level, the proof differs from that of the 2-way part in two aspects:
\begin{enumerate}
\item The underlying measure that is being used.
\item The partition of the $(a,b)$-plane into regions, which is no longer symmetric with respect to Alice and Bob as can be seen in Figure~\ref{fig:division-1}.
\end{enumerate}

\noindent We introduce a new measure $m^{\oneway}(f)$ as follows (this is the 1-way analog of Proposition~\ref{prop:measure_2way_total}).

\begin{defn}[Measure $m^{\oneway}(f)$ for total functions] \label{def:1way_measure}
Given a total permutation-invariant function $f : \bit^n \times \bit^n \to \bit$, and integers $a, b$, s.t. $0 \le a, b \le n$, let $h_{a,b} : \set{0, 1, \cdots, n} \to \set{0,1,?}$ be the function given by $h_{a,b}(d) = f(\bx, \by)$ if there exist $\bx$, $\by$ with $|\bx| = a$, $|\by| = b$, $\Delta(\bx, \by) = d$ and $?$ otherwise. (Note. by permutation invariance of $f$, $h_{a,b}$ is well-defined.) Let $\calJ(h_{a,b})$ be the set of {\em jumps} in $h_{a,b}$, defined as follows,
$$\calJ(h_{a,b}) \quad \defeq \quad \setdef{c}{\inmat{h_{a,b}(c-1) \ne h_{a,b}(c+1) \\ h_{a,b}(c-g), \ h_{a,b}(c+g) \in \bit}}$$
Then, we define $m^{\oneway}(f)$ as follows.
$$m^{\oneway}(f) \quad \defeq \quad \frac{1}{C} \cdot \max_{\substack{a, b \in [n] \\ (c,g) \in \calJ(h_{a,b})}} \max\set{\min\set{a,c,n-a,n-c}, \log\inparen{\min \set{c, n-c}}}$$
where $C$ is a suitably large constant (which does not depend on $n$).
\end{defn}

\noindent We point out that the only difference between Definition~\ref{def:1way_measure} and Proposition~\ref{prop:measure_2way_total} is that the term $\min\inbrace{a,b,c,n-a,n-b,n-c}$ in Proposition~\ref{prop:measure_2way_total} is replaced by $\min\inbrace{a,c,n-a,n-c}$ in Definition~\ref{def:1way_measure}. In particular, Definition~\ref{def:1way_measure} is not symmetric with respect to Alice and Bob.

\noindent We will now prove the following theorem, which immediately implies the 1-way part of Theorem~\ref{thm:main-2+3}.

\begin{theorem}\label{thm:isr_thm_1_way}
Let $f : \set{0, 1}^n \times \set{0, 1}^n \rightarrow \set{0, 1}$ be any (total) permutation-invariant function. Then,
$$m^{\oneway}(f) \le \R^{\oneway}(f) \le \isr^{\oneway}(f) \le \Otilde(m^{\oneway}(f)^3) + \R^{\oneway}(f) + O(\log \log n)$$
\end{theorem}

We will need the following lemma to show that the measure $m^{\oneway}(f)$ is a lower bound on $\R^{\oneway}(f)$. This is analogous to Lemma~\ref{lem:main_low_bds_ghd}, which was used to show that $m(f)$ is a lower bound on $\R(f)$ and in particular, $\IC(f)$.

\begin{lem}\label{lem:1way_GHD_LB}
For all $n, a, b, c, g$ for which $\eGHD^n_{a,b,c,g}$ is a meaningful problem, the following lower bounds hold,
$$\R^{\oneway}(\eGHD^n_{a,b,c,g}) \geq \frac{1}{C} \cdot \inparen{\frac{\min\set{a, n-a, c, n-c}}{g}}$$
$$\R^{\oneway}(\eGHD^n_{a,b,c,g}) \geq \frac{1}{C} \cdot \inparen{\log\inparen{\frac{\min \set{c, n-c}}{g}}}$$
\end{lem}

We defer the proof of Lemma~\ref{lem:1way_GHD_LB} to Section~\ref{sec:1way_GHD_low_bds}. For now, we will use this lemma to prove Theorem~\ref{thm:isr_thm_1_way}.

\begin{proof}
Any protocol to compute $f$ also computes $\eGHD^n_{a,b,c,1}$ for any $a$, $b$ and any jump $c \in \calJ(h_{a,b})$. Consider the choice of $a$, $b$ and a jump $c \in \calJ(h_{a,b})$ such that the lower bound obtained on $\IC(\eGHD^n_{a,b,c,1})$ through Lemma~\ref{lem:1way_GHD_LB} is maximized, which by definition is equal to $m^{\oneway}(f)$ (after choosing an appropriate $C$ in the definition of $m^{\oneway}(f)$). Thus, we have $m^{\oneway}(f) \le \R^{\oneway}(\eGHD^n_{a,b,c,g}) \le \R^{\oneway}(f)$.

Let $k \defeq m^{\oneway}(f)$. The main part of the proof is to show that $\isr^{\oneway}(f) \le \Otilde(k^3) + \R^{\oneway} + O(\log \log n)$. We first divide the input space into a constant number of regions, such that Alice can send $O(\log{k})$ bits to Bob with which he can decide in which of the regions does their combined input lie (with high probability). Thus, once we break down the input space into these regions, it will suffice to give $1$-way protocols with small ISR-CC for computing the function over each of these regions; as Alice can then send the 1-way messages corresponding to all of these regions, and Bob will first determine in which region their combined input lies and then use the corresponding messages of Alice to compute the function value.

Suppose that we have a function $f$ with $m^{\oneway}(f) = k$. Let $a = |\bx|$ and $b = |\by|$. We partition the $(a,b)$-plane into $4$ regions based on the values of $a$ and $b$ as follows. First let $A = \min\set{a, n-a}$ and $B = \min\set{b, n-b}$. Then the regions are given by,

\begin{enumerate}
\item[(I)] $A \le Ck$ and $B \le 2^{Ck}$
\item[(II)] $A \le Ck$ and $B > 2^{Ck}$
\item[(III)] $A > Ck$ and $|A-B| < Ck$
\item[(IV)] $A > Ck$ and $|A-B| \geq Ck$
\end{enumerate}

\noindent where $C$ comes from Definition~\ref{def:1way_measure}. Note that the regions (I), (II), (III) and (IV) form a partition of the $(a,b)$ plane. This partition is shown pictorially in Figure~\ref{fig:division-1}.

First, note that if $|\bx| > n/2$, then Alice can flip all her input bits and convey that she did so to Bob using one bit of communication. Similarly, if $|\by| > n/2$, then Bob can flip all his input bits and convey that he did so to Alice using one bit of communication. After these flipping operations, Alice and Bob will look at the appropriately modified version of $f$ based on who all flipped their input. Note that flipping all the bits of Alice and/or Bob preserves the permutation-invariance of the function. We will henceforth assume w.l.o.g. that $a = |\bx| \le n/2$ and $b = |\by| \le n/2$. [This is also the reason that the regions described above are in terms of $A = \min\set{a, n-a}$ and $B = \min\set{b, n-b}$ and $A, B \le n/2$.]

We now show that determining the region to which the input pair $(\bx, \by)$ belongs can be done using $O(\log{k})$ bits of (ISR) communication from Alice to Bob. First, to distinguish between \{(I), (II)\} and \{(III), (IV)\}, Alice needs to send one bit to Bob indicating whether $a \le 100k$. Moreover, in the case of \{(I), (II)\}, Bob can easily differentiate between (I) and (II) because he knows the value of $b$. To distinguish between regions (III) and (IV), Alice and Bob can first check whether $|a-b| < 100k$ by setting up an instance of {\sc Sparse-Indexing}. Namely, Alice will translate the value $a$ into a string $s_a$ where $s_a(i) = 1$ iff $i=a$. And Bob will translate $b$ into a string $s_b$ such that $s_b(i) = 1$ iff $b-Ck < i < b+Ck$. This is an instance of {\sc Sparse-Indexing} which can be solved with $O(\log k)$ bits of ISR-CC, by Corollary~\ref{cor:isr_sparse_ind_cor}.

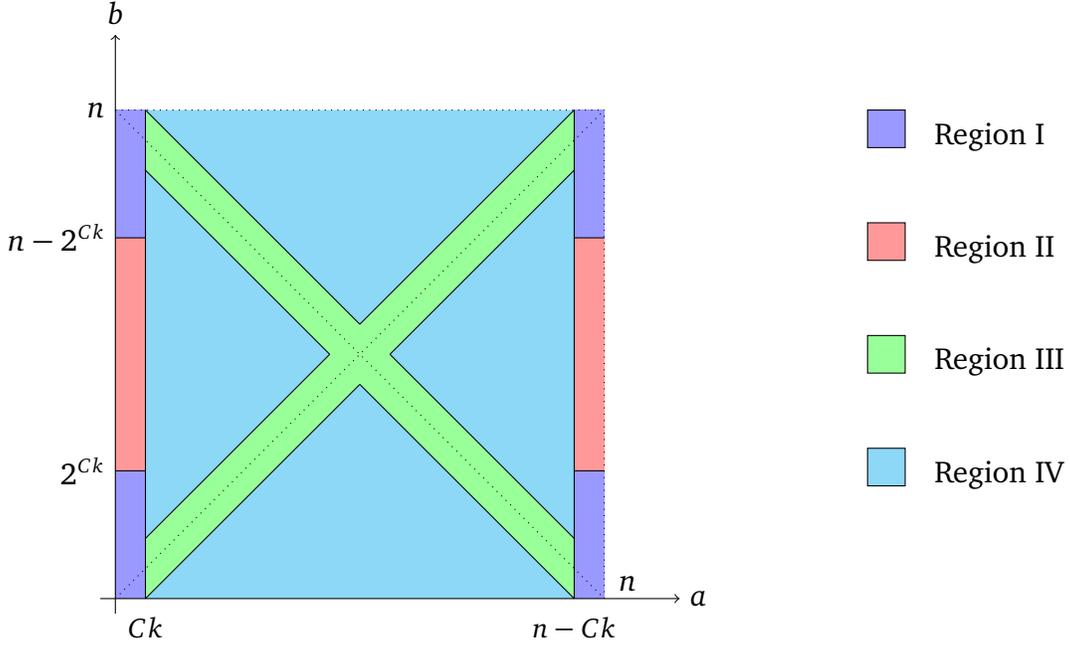
\begin{figure}
\begin{center}
\begin{tikzpicture}
\def \n {6.5};
\def \ex {1};
\def \smex {0.1};
\def \k {0.4};
\def \expk {1.7};

\def \index {\n+4};
\def \len {0.5};
\def \hgt {1.5};

\def \bgt{40}
\def \colA {blue!\bgt};
\def \colB {red!\bgt};
\def \colC {green!\bgt};
\def \colD {cyan!\bgt};

\coordinate (A) at (0,0);
\coordinate (B) at (\n,0);
\coordinate (C) at (\n,\n);
\coordinate (D) at (0,\n);

\coordinate (K0) at (\k,0);
\coordinate (K1) at (\k,\n);
\coordinate (K01) at (\k,2*\k);
\coordinate (K10) at (\k,\n-2*\k);
\coordinate (K2) at (\n-\k,0);
\coordinate (K3) at (\n-\k,\n);
\coordinate (K02) at (\n-\k,2*\k);
\coordinate (K20) at (\n-\k,\n-2*\k);

\coordinate (E0) at (0,\expk);
\coordinate (E1) at (\k,\expk);
\coordinate (E00) at (0,\n-\expk);
\coordinate (E10) at (\k,\n-\expk);
\coordinate (E2) at (\n-\k, \n-\expk);
\coordinate (E3) at (\n, \n-\expk);
\coordinate (E20) at (\n-\k, \expk);
\coordinate (E30) at (\n, \expk);

\coordinate (M1) at (\n/2,\n/2-\k);
\coordinate (M2) at (\n/2+\k,\n/2);
\coordinate (M3) at (\n/2,\n/2+\k);
\coordinate (M4) at (\n/2-\k,\n/2);

\draw[draw=none, fill=\colA] (A) -- (K0) -- (E1) -- (E0) -- cycle;
\draw[draw=none, fill=\colA] (C) -- (K3) -- (E2) -- (E3) -- cycle;
\draw[draw=none, fill=\colA] (D) -- (K1) -- (E10) -- (E00) -- cycle;
\draw[draw=none, fill=\colA] (B) -- (K2) -- (E20) -- (E30) -- cycle;

\draw[draw=none, fill=\colB] (E0) -- (E1) -- (E10) -- (E00) -- cycle;
\draw[draw=none, fill=\colB] (E3) -- (E2) -- (E20) -- (E30) -- cycle;

\draw[draw=none, fill=\colC] (K0) -- (M1) -- (K2) -- (K02) -- (M2) -- (K20) -- (K3) -- (M3) -- (K1) -- (K10) -- (M4) -- (K01) -- cycle;

\draw[draw=none, fill=\colD] (K0) -- (M1) -- (K2) -- cycle;
\draw[draw=none, fill=\colD] (K02) -- (M2) -- (K20) -- cycle;
\draw[draw=none, fill=\colD] (K3) -- (M3) -- (K1) -- cycle;
\draw[draw=none, fill=\colD] (K10) -- (M4) -- (K01) -- cycle;

\node[right] (a) at (\n+\ex,0) {$a$};
\node[above] (b) at (0,\n+\ex) {$b$};

\draw[->] (-\ex/5,0) -- (a);
\draw[->] (0,-\ex/5) -- (b);

\draw (K0) -- (K1);
\draw (K2) -- (K3);

\draw (E0) -- (E1);
\draw (E00) -- (E10);
\draw (E2) -- (E3);
\draw (E20) -- (E30);

\draw (K0) -- (M1);
\draw (M2) -- (K20);
\draw (K01) -- (M4);
\draw (M3) -- (K3);
\draw (K10) -- (M4);
\draw (M1) -- (K2);
\draw (K1) -- (M3);
\draw (M2) -- (K02);

\draw[dotted] (A) -- (C);
\draw[dotted] (D) -- (B);
\draw[dotted] (D) -- (C);
\draw[dotted] (B) -- (C);

\node[below] at (\k,-\smex) {\small $Ck$};
\node[below] at (\n-\k,-\smex) {\small $n-Ck$};
\node[right] at (\n+0.5*\smex,2*\smex) {$n$};
\node[left] at (0,\n) {$n$};
\node[left] at (0,\expk) {$2^{Ck}$};
\node[left] at (0,\n-\expk) {$n-2^{Ck}$};

\draw[fill=\colD] (\index, \hgt) rectangle (\index-\len, \hgt+\len);
\draw[fill=\colC] (\index, 2*\hgt) rectangle (\index-\len, 2*\hgt+\len);
\draw[fill=\colB] (\index, 3*\hgt) rectangle (\index-\len, 3*\hgt+\len);
\draw[fill=\colA] (\index, 4*\hgt) rectangle (\index-\len, 4*\hgt+\len);

\node[right] at (\index+\len/2, \hgt+\len*0.3) {Region IV};
\node[right] at (\index+\len/2, 2*\hgt+\len*0.3) {Region III};
\node[right] at (\index+\len/2, 3*\hgt+\len*0.3) {Region II};
\node[right] at (\index+\len/2, 4*\hgt+\len*0.3) {Region I};

\end{tikzpicture}
\caption{Division of the $(a,b)$ plane into four regions in the $1$-way setting}
\label{fig:division-1}
\end{center}
\end{figure}


We now show how to compute the value of the function $f$ in each of the $4$ individual regions (I), (II), (III) and (IV) using 1-way ISR-CC of at most $\Otilde(k^3) + \R^{\oneway}(f) + O(\log \log n)$ bits. Since $f$ is a permutation-invariant function, we use the following $2$ interchangeable representations of $f$ using Observation~\ref{obs:interchang_rep},
$$f(\bx, \by) = g(|\bx|, |\by|, |\bx \land \by|) = h(|\bx|, |\by|, \Delta(\bx, \by))$$

\begin{enumerate}

\item[(I)] (Main idea: {\sc Small-Set-Intersection}) We have that $a \le Ck$ and $b \le 2^{Ck}$. Alice can first send the value of $a = |\bx|$ to Bob. They can then apply the protocol from Lemma~\ref{le:ssi_isr_prot} in order to compute $|\bx \land \by|$ using $O(a\log(ab)) = O(k^2)$ bits of $1$-way ISR communication from Alice to Bob. Hence, Bob can determine $|\bx \land \by|$ correctly with high probability, and hence deduce $g(|\bx|, |\by|, |\bx \land \by|) = f(\bx, \by)$.

\item[(II)] (Main idea: {\em No dependence on $\Delta(\bx, \by)$}) We have that $a \le Ck$ and $b > 2^{Ck}$. In this case, the definition of our measure $m^{\oneway}(f)$ implies that for this range of values of $a$ and $b$, the function $h$ cannot depend on $\Delta(\bx, \by)$ (because in this case $\Delta(\bx, \by) \ge (b-a)$). Since $h$ depends only on $|\bx|$ and $|\by|$, Alice can simply send the value of $a$ (which takes only $O(\log k)$ bits), with which Bob can compute $h(a, b, c)$ for any valid $c$, that is, $h(a,b,c) \ne \ ?$.

\item[(III)] (Main idea: {\sc Small-Hamming-Distance}) We have that that $|a-b| < Ck$. Then, Alice sends the value of $a~(\mod 2Ck)$ to Bob, requiring $O(k)$ 1-way communication. Since Bob knows $b$, he can figure out the exact value of $a$. Next, they need to determine the Hamming distance $\Delta(\bx, \by)$. The definition of our measure $m^{\oneway}(f)$ (along with the fact that $k:=m^{\oneway}(f)$) implies that if $h(a,b,c-1) \ne h(a,b,c+1)$, then $c$ must be either $\le Ck$ or $\geq n-Ck$. That is, given $a$ and $b$, $h(a, b, c)$ must equal a constant for all valid $c$ such that $Ck < c < n - Ck$.

Since Bob knows both $a$ and $b$ exactly, Alice and Bob can run the 1-way ISR-protocol for $\HD^n_i$ (from Lemma~\ref{le:main_isr_up_bd_k_HD}) for every $i < Ck$ and $i > n-Ck$. This requires $\Otilde(k^3)$ communication.\footnote{The ISR-protocol for $\HD^n_i$ takes $O(i^2) \le O(k^2)$ communication each. But we can only afford to make error with probability $O(1/k)$, and thus for sake of amplification, the overall communication is $O(k^3 \log k)$.} If the Hamming distance is either $\le Ck$ or $\geq n-Ck$, then Bob can determine it exactly and output $h(a,b,c)$. If the Hamming distance is not in this range, then the Hamming distance does not influence the value of the function, in which case Bob can output $h(a, b, c)$ for any valid $c$ such that $Ck < c < n - Ck$.

\item[(IV)] (Main idea: {\em Newman's theorem} on $(a,b)$) We claim that for this range of values of $a$ and $b$, the function $h(a, b, c)$ cannot depend on $c$. Suppose for the sake of contradiction that the function depends on $c$. Hence, there exists a value of $c$ for which $h(a, b, c-1) \ne h(a, b, c+1)$. Since $|a - b| \ge Ck$, we have that $c \ge Ck$. And since $|n-a-b| \geq Ck$, we also have that $c \le \min\set{a+b, 2n-a-b} \le n - Ck$. Thus, we get that $Ck \le c \le n-Ck$, which contradicts that $m^{\oneway}(f) = k$ (see Definition~\ref{def:1way_measure}).

Since $f(\bx, \by)$ only depends on $|\bx|$ and $|\by|$ in this region, we have converted the problem which depended on inputs of size $n$, into a problem which depends on inputs of size $O(\log{n})$ only. Since the original problem had a 1-way PSR-protocol with $\R^{\oneway}(f)$ bits of communication, applying Newman's theorem (Theorem~\ref{thm:newman}), we conclude that, with private randomness itself, the problem can be solved using $\R^{\oneway}(f) + O(\log \log n)$ bits of 1-way communication.\qedhere
\end{enumerate}

\end{proof}

\begin{note}
We point that the proof of Theorem~\ref{thm:isr_thm_1_way} shows, more strongly, that for any function $G(.)$, the following two statements are equivalent:
\begin{enumerate}
\item[(1)] Any total function $f$ with $\R^{\oneway}(f) = k$ has $\isr^{\oneway}(f) \le \poly(k) + G(n)$
\item[(2)] Any total permutation-invariant function $f$ with $\R^{\oneway}(f) = k$ has $\isr^{\oneway}(f) \le \poly(k) + G(O(\log n))$.
\end{enumerate}
Newman's theorem (Theorem~\ref{thm:newman}) implies that (1) holds for $G(n) = O(\log{n})$, and hence we have that (2) also holds for $G(n) = O(\log{n})$; thereby yielding the bound implied by Theorem~\ref{thm:isr_thm_1_way}. On the other hand, improving the $O(\log \log n)$ term in Theorem~\ref{thm:isr_thm_1_way} to, say $o(\log \log n)$ will imply that for all total functions we can have $G(n)$ in (1) to be $o(\log n)$. We note that currently such a statement is unknown.
\end{note}
\fi


\subsection{1-way CC lower bounds on Gap Hamming Distance} \label{sec:1way_GHD_low_bds}
In this section, we prove lower bounds on 1-way randomized communication complexity of {\sc Gap-Hamming-Distance} (Lemma~\ref{lem:1way_GHD_LB}).

\ifnum\focs=1
The proof of Lemma~\ref{le:1way_HD_LB} appears in the full version. We point out that part $2$ of the lemma is used in the proof of the $1$-way part of Theorem~\ref{thm:main-2+3}.
\fi

\noindent We will prove Lemma~\ref{lem:1way_GHD_LB} by getting certain reductions from {\sc Sparse-Indexing} (namely Proposition~\ref{lem:1way_basic_ludisj_lb}). Similar to our approach in Section~\ref{sec:GHD_results}, we first prove lower bounds on 1-way PSR-CC of {\sc Set-Inclusion} (Definition~\ref{defn:set_inclusion}). We do this by obtaining a reduction from {\sc Sparse-Indexing}.

\begin{proposition}[{\sc Set-Inclusion} 1-way PSR-CC lower bound] \label{prop:lopsided_set_inc_low_bd}
For all $t, w \in \bbN$, $$\R^{\oneway}(\setinc^{t+w}_{t, t+w-1}) \ge \Omega(\min(t, w))$$
\end{proposition}
\begin{proof}
We know that $\R^{\oneway}(\SI^{t+w}_t) \ge \Omega(\min(t,w))$ from Proposition \ref{prop:ludisj_low_bd}. Note that $\SI^{t+w}_t$ is same as the problem of $\eGHD^{t+w}_{t,1,t,1}$. If we instead think of Bob's input as complemented, we get that solving $\eGHD^{t+w}_{t,t+w-1,w,1}$ is equivalent to solving $\eGHD^{t+w}_{t,1,t,1}$, which is same as $\setinc^{t+w}_{t,t+w-1}$. Thus, we conclude that $\R^{\oneway}(\setinc^{t+w}_{t,t+w-1}) \ge \Omega(\min(t,w))$.
\end{proof}

\noindent We now state and prove a technical lemma that will help us prove Lemma~\ref{lem:1way_GHD_LB}. Note that this is a 1-way analogue of Lemma~\ref{le:first_low_bds_ghd}.
\begin{lem}\label{le:first_1way_low_bds_ghd}
Let $a, b \le n/2$. Then, the following lower bounds hold,
\begin{enumerate}
\item[(i)] $\R^{\oneway}(\eGHD^n_{a,b,c,g}) \ge \Omega\inparen{\min\inbrace{\frac{c-b+a}{g}, \frac{n-c}{g}}}$
\item[(ii)] $\R^{\oneway}(\eGHD^n_{a,b,c,g}) \ge \Omega\inparen{\min\inbrace{\frac{a+b-c}{g}, \frac{c+b-a}{g}}}$
\item[(iii)] $\R^{\oneway}(\eGHD^n_{a,b,c,g}) \ge \Omega\inparen{\min\inbrace{\log \inparen{\frac{c}{g}}, \log \inparen{\frac{n-c}{g}}}}$
\end{enumerate}
\end{lem}
\begin{proof}
We first note that part (iii) follows trivially from Lemma~\ref{le:first_low_bds_ghd} because $\IC(f) \le \R(f) \le \R^{\oneway}(f)$. But we require slightly different proofs for parts (i) and (ii). The main technical difference in these proofs and the corresponding ones in Lemma~\ref{le:first_low_bds_ghd} is that now we cannot assume that $a \le b \le n/2$. We note that this assumption of $a \le b$ was extremely crucial for Lemma~\ref{le:first_low_bds_ghd}, without which parts (i) and (ii) would not have been possible (even though they look exactly same in form!). The reason why we are able to succeed here without making the assumption because we derive reductions from a much ``simpler'' problem, but which is still hard in the 1-way model, namely the {\sc Sparse-Indexing} problem.\\

\noindent {\bf Proof of (i).} We obtain a reduction from $\SI^{2t}_t$ for which we know from Proposition~\ref{prop:ludisj_low_bd} that $\R^{\oneway}(\SI^{2t}_t) = \Omega(t)$. Recall that $\SI^{2t}_t$ is same as $\eGHD^{2t}_{t,1,t,1}$. We repeat the instance $g$ times to get an instance of $\eGHD^{2gt}_{gt, g, gt, g}$. Now, we need to append $(a-gt)$ 1's to $\bx$ and $(b-g)$ 1's to $\by$. This will increase the Hamming distance by a fixed amount which is at least $|a-b-gt+g|$ and at most $b+a-gt-g$. Also, the number of inputs we need to add is at least $((a-gt)+(b-g)+(c-gt))/2$. Thus, we can get a reduction to $\overline{\GHD}^n_{a,b,c,g}$ if and only if,
$$|a-b-gt+g| \le c - gt \le b+a-gt-g$$
$$n \ge 2gt + \frac{(a-gt)+(b-g)+(c-gt)}{2}$$

The left condition on $c$ gives us that $2gt \le c + a - b + g$ and $b-a+g \le c$ (which is always true). The right condition on $c$ gives $c \le b+a-g$ (which is always true). The condition on $n$ gives that $gt \le n-(a+b+c-g)/2$, which is equivalent to $$t \le \frac{n-a-b}{2g} + \frac{n-c-g}{2g}$$
Thus, we can have the reduction work by choosing $t$ to be
$$t = \min\inbrace{\frac{c-b+a+g}{2g},\frac{n-c-g}{2g}} = \Omega\inparen{\min\inbrace{\frac{c-b+a}{g}, \frac{n-c}{g}}}$$
(since $n \ge a+b$) and thus we obtain
$$\R^{\oneway}(\eGHD^n_{a,b,c,g}) \ge \R^{\oneway}(\SI^{2t}_t) \ge \Omega\inparen{\min\inbrace{\frac{c-b+a}{g}, \frac{n-c}{g}}}$$

\noindent {\bf Proof of (ii).} We obtain a reduction from $\setinc^{m}_{t,t+w-1}$ (where $m=t+w$) for which we know from Proposition~\ref{prop:lopsided_set_inc_low_bd} that $\R^{\oneway}(\SI^{m}_{t,t+w-1}) = \Omega(\min\set{t,w})$. Recall that $\setinc^{m}_{t,t+w-1}$ is same as $\eGHD^{m}_{t,t+w-1,w,1}$. Given an instance of $\overline{\GHD}^{m}_{t,t+w-1,w,1}$, we first repeat the instance $g$ times to get an instance of $\overline{\GHD}^{gm}_{gt,g(t+w-1),gw,g}$. Now, we need to append $(a-gt)$ 1's to $\bx$ and $(b-gt-gw+g)$ 1's to $\by$. This will increase the Hamming distance by a fixed amount which is at least $|b-a-gw+g|$ and at most $(b-gt-gw+g)+(a-gt)$. Also, the number of inputs we need to add is at least $((a-gt)+(b-gt-gw+g)+(c-gw))/2$. Thus, we can get a reduction to $\eGHD^n_{a,b,c,g}$ if and only if,
\begin{eqnarray*}
|b-a-gw+g| \le c - gw \le b+a-2gt-gw+g \\
n \ge gt+gw + \frac{(b-gt-gw+g)+(a-gt)+(c-gw)}{2}
\end{eqnarray*}

\noindent The left constraint on $c$ requires $c \ge \max\set{b-a+g, a+2gw-b-g}$. We know that $c \ge b-a+g$, so the only real constraint is $c \ge 2gw - (b-a) + g$, which gives us that,
$$w \le \frac{c+b-a-g}{2g}$$
The right constraint on $c$ requires $c \le b+a-2gt+g$, which gives us that,
$$t \le \frac{a+b-c+g}{2g}$$
Suppose we choose $t$ to be $\frac{a+b-c+g}{2g}$. Then the constraint on $n$ is giving us that,
$$n \ge gt + \frac{a+b+c-g}{2} = \frac{a+b-c+g}{2} + \frac{a+b+c-g}{2} = a+b$$
We already assumed that $a, b \le n/2$, and hence this is always true.

\noindent Thus, we can choose $t = \frac{a+b-c+g}{2g}$ and $w = \frac{c+b-a-g}{2g}$, and from Proposition \ref{prop:lopsided_set_inc_low_bd}, we get,
$$\R^{\oneway}(\eGHD^n_{a,b,c,g}) \ge \R^{\oneway}(\setinc^{t+w}_{t,w}) \ge \min(\set{t,w}) \ge \Omega\inparen{\min\inbrace{\frac{a+b-c}{g}, \frac{c+b-a}{g}}}$$
\end{proof}

\noindent We are now finally able to prove Lemma~\ref{lem:1way_GHD_LB}.

\begin{proofof}{Lemma~\ref{lem:1way_GHD_LB}}
Assume for now that $a, b \le n/2$. From parts (i) and (ii) of Lemma~\ref{le:first_1way_low_bds_ghd}, we know the following,
\begin{eqnarray*}
\R^{\oneway}(\eGHD^n_{a,b,c,g}) &\ge& \Omega\inparen{\min\inbrace{\frac{c-b+a}{g}, \frac{n-c}{g}}}\\
\R^{\oneway}(\eGHD^n_{a,b,c,g}) &\ge& \Omega\inparen{\min\inbrace{\frac{a+b-c}{g}, \frac{c+b-a}{g}}}
\end{eqnarray*}

\noindent Adding these up, we get that,
$$\R^{\oneway}(\eGHD^n_{a,b,c,g}) \ge \Omega\inparen{\min\inbrace{\frac{c-b+a}{g}, \frac{n-c}{g}} + \min\inbrace{\frac{a+b-c}{g}, \frac{c+b-a}{g}}}$$

\noindent Since $\min\set{A,B} + \min\set{C,D} = \min\set{A+C, A+D, B+C, B+D}$, we get that,
$$\R^{\oneway}(\eGHD^n_{a,b,c,g}) \ge \Omega\inparen{\min\inbrace{\frac{2a}{g}, \frac{2c}{g}, \frac{n+a+b-2c}{g}, \frac{n+b-a}{g}}}$$
For the last two terms, note that, $n+a+b-2c \ge n-c$ (since $a+b \ge c$) and $n+b-a \ge n-a \ge a$ (since $n/2 \ge a$). Thus, overall we get,
$$\R^{\oneway}(\eGHD^n_{a,b,c,g}) \ge \Omega\inparen{\min\inbrace{\frac{a}{g}, \frac{c}{g}, \frac{n-c}{g}}}$$

\noindent Note that this was assuming $a, b \le n/2$. In general, we get,
$$\R^{\oneway}(\eGHD^n_{a,b,c,g}) \ge \Omega\inparen{\frac{\min\inbrace{a,c,n-a,n-c}}{g}}$$
[We get the $(n-a)$ term because we might have flipped Alice's input to make sure $a \le n/2$ But unlike in the proof of Lemma~\ref{lem:main_low_bds_ghd}, we don't get $b$ or $(n-b)$ in the above lower bound because while restricting to $a,b \le n/2$, we never flipped the role of Alice and Bob.]

\noindent The second part follows trivially from the corresponding part of Lemma~\ref{lem:main_low_bds_ghd}, since $\IC(f) \le \R(f) \le \R^{\oneway}(f)$ for any $f$.

\noindent We choose $C$ to be a large enough constant, so that the desired lower bounds hold.
\end{proofof}

\ifnum\focs=1

On a high level, the main difference in the proof of the one-way version of Theorem~\ref{thm:main-2+3} is that in this case the partition of the regions is \emph{asymmetric} with respect to Alice and Bob as shown in Figure~\ref{fig:division-1}. For the full details, we refer the reader to the full version.
\fi


\section{Conclusion} \label{sec:conclusion}

In this work, we initiated the study of the communication complexity of permutation-invariant functions. We gave a coarse characterization of their information complexity and communication complexity (Theorem~\ref{thm:main-1}). We also showed for total permutation-invariant functions that the communication complexity with imperfectly shared randomness is not much larger than the communication complexity with perfectly shared randomness (Theorem~\ref{thm:main-2+3}). Our work points to several possible future directions.

\begin{itemize}
\item {[{\bf Generalized Permutation-Invariance}]} Is it possible to generalize our results for any larger class of functions? One candidate might be classes of functions that are invariant under natural subgroups of the permutation group $S_n$, or more generally any group of actions on the input spaces of Alice and Bob. For example, once choice of a subgroup of permutations is the set of all permutations on $[n]$, that map $[n/2]$ to $[n/2]$ and map $[n] \setminus [n/2]$ to $[n] \setminus [n/2]$. Or more generally, the subgroup of $\ell$-part permutation-symmetric functions, which consists of functions $f$ for which there is a partition $I = \set{I_1, \dots , I_{\ell}}$ of $[n]$ such that $f$ is invariant under any permutation $\pi \in S_n$ where $\pi(I_i) = I_i$ for every $i \in \{1, \dots , \ell\}$.

\item {[{\bf Permutation-Invariance over higher alphabets}]} Another interesting question is to generalize our results to larger alphabets, i.e., to permutation-invariant functions of the form $f:\mathcal{X}^n \times \mathcal{Y}^n \to R$ where $\mathcal{X}$, $\mathcal{Y}$ and $R$ are not necessarily binary sets.

\item {[{\bf Tight lower bounds for Gap-Hamming-Distance}]} What are the tightest lower bounds on $\GHD^n_{a,b,c,g}$ for all choices of parameters, $a$, $b$, $c$, $g$? Our lower bounds on $\GHD^n_{a,b,c,g}$ are not tight for all choices of parameters $a$, $b$, $c$ and $g$. For example, when $a = b = c = n/2$ and $g = \sqrt{n}$, our lower bound in Lemma~\ref{lem:main_low_bds_ghd} only implies $\IC(\GHD^{n}_{a,b,c,g}) \ge \Omega(\sqrt{n})$. By slightly modifying the proof techniques in \cite{chakrabarti2012optimal, vidick2011concentration, sherstov2012communication}, one can obtain that $\R(\GHD^{n}_{a,b,c,g}) = \Omega(n)$. Sherstov's proof \cite{sherstov2012communication} is based on the {\em corruption bound}. Combining this with the recent result due to \cite{kerenidis2012lower} which showed (by studying a relaxed version of the partition bound of \cite{jain2010partition}) that many known lower bound methods for randomized communication complexity -- including the corruption bound -- are also lower bounds for the information complexity. This implies that $\IC(\GHD^{n}_{a,b,c,g}) = \Omega(n)$ for $a = b = c = n/2$ and $g = \sqrt{n}$.

\item {[{\bf Hierarchy within ISR}]} The work of \cite{CGMS_ISR} shows an exponential separation between $\isr_{\rho}(f)$ and $\R(f)$. However, it is unclear if some {\em strong separation} could be shown between $\isr_{\rho}(f)$ and $\isr_{\rho'}(f)$ for some function $f$ (where $\rho < \rho' < 1$).

\item {[{\bf Limits of separation in \cite{CGMS_ISR}}]} Canonne et al showed that for some {\em unspecified} $k$, there is a {\em partial} permutation-invariant function with communication at most $k$ under perfect sharing of randomness, but with communication at least $\Omega(2^k)$ under imperfect sharing of randomness. Can their separation be made to hold for $k = \Theta(\log\log{n})$? Answering this question would shed some light on the possibility of proving an analogue of Theorem~\ref{thm:main-2+3} for partial permutation-invariant functions.
\end{itemize}


\bibliographystyle{alpha}
\newcommand{\etalchar}[1]{$^{#1}$}


\end{document}